\documentclass[10pt,twocolumn,aps,prd,preprintnumbers,showpacs,superscriptaddress,nofootinbib,amsmath,amssymb,floats,floatfix,showkeys,notitlepage,longbibliography]{revtex4-2}
\usepackage{orcidlink}
\usepackage{comment}
\usepackage{lipsum}
\usepackage{graphicx}
\usepackage{subfigure}
\usepackage{hyperref}
\usepackage{palatino}
\usepackage{sans}
\usepackage{hyperref}
\hypersetup{colorlinks=true,linkcolor=blue,urlcolor=blue,citecolor=blue}
\usepackage[toc,page]{appendix}
\usepackage[normalem]{ulem}
\usepackage{adjustbox}
\usepackage{latexsym}
\usepackage{amsmath}
\usepackage{amssymb}
\usepackage{amsfonts}
\usepackage{mathrsfs}
\usepackage{dcolumn}
\usepackage{bm}
\usepackage{tikz}
\usetikzlibrary{decorations.pathmorphing}
\usepackage{bigints}
\usepackage{array,tabularx,multirow,booktabs}
\usepackage[tracking=true]{microtype}
\usepackage{soul} 
\SetTracking{}{500}
\SetTracking{encoding={*}, shape=sc}{40}
\UseRawInputEncoding 
\allowdisplaybreaks

\usepackage[utf8]{inputenc}
\usepackage{xcolor} 

\begin{document} \sloppy

\title{Acceleration radiation from vibrating atoms in Schwarzschild spacetime}

\author{Reggie C. Pantig \orcidlink{0000-0002-3101-8591}} 
\email{rcpantig@mapua.edu.ph}
\affiliation{Physics Department, School of Foundational Studies and Education, Map\'ua University, 658 Muralla St., Intramuros, Manila 1002, Philippines.}

\author{Ali \"Ovg\"un \orcidlink{0000-0002-9889-342X}}
\email{ali.ovgun@emu.edu.tr}
\affiliation{Physics Department, Eastern Mediterranean University, Famagusta, 99628 North
Cyprus via Mersin 10, Turkiye.}

\author{Syed Masood}
\email{masood@westlake.edu.cn}
\email{quantummind137@gmail.com}
\affiliation{Department of Physics, School of Science and Research Center for Industries of the Future, Westlake University, Hangzhou $310030$, P. R. China}
\affiliation{Institute of Natural Sciences, Westlake Institute for Advanced Study, Hangzhou $310024$, P. R. China.}

\author{Li-Gang Wang}
\email{lgwang@zju.edu.cn}
\affiliation{School of Physics, Zhejiang University, Hangzhou 310027, P.R. China.}

\begin{abstract}
Motivated by the work of Scully \textit{et al.} [ \textcolor{blue}{Proc. Nat. Acad. Sci. 115, 8131 (2018)}] and Dolan \textit{et al.}[ \textcolor{blue}{New J. Phys. 22, 033026 (2020)}], we study the acceleration radiation from a two-level Unruh-DeWitt detector that undergoes small-amplitude radial oscillations at fixed mean radius $R_0$ outside a Schwarzschild black hole. The massless scalar field is quantized in the Boulware vacuum to isolate curvature-modulated acceleration effects without a thermal Hawking background. Working in a (1+1) radial reduction and using first-order time-dependent perturbation, we evaluate the period-averaged transition rate (or the Floquet transition rate). The resulting particle emission spectrum exhibits a thermal Bose-Einstein-type profile with periodic trajectory yielding a Floquet resonance condition $n\Omega > \omega_0$ and a closed-form expression for the Floquet transition rate $\overline{P}_n$, which reduces to the flat Minkowski spacetime result as $R_0\to\infty$, in agreement with  Near the horizon, $f(R_0)<1$ enhances the effective Bessel argument by $1/\sqrt{f(R_0)}$, providing a simple analytic demonstration of curvature/redshift amplification of acceleration radiation. In particular, the spectrum weighted by the Bessel function becomes ill-defined near the black hole horizon as $R_{0}\rightarrow 2M$, possibly manifesting the well-known pathological behavior ofthe  Boulware vacuum state. We discuss the regime of validity (small amplitude, $R_0$ away from the horizon) and outline the extensions to (3+1) dimensions, including density-of-states and greybody factors, and to alternative vacuum choices. Our results offer an analytically tractable link between flat-space vibrating atom proposals and black-hole spacetimes.
\end{abstract}


\maketitle

\section{Introduction} \label{sec1}
In quantum field theory on curved spacetimes, the link between gravitational fields and quantum vacuum fluctuations gives rise to remarkable phenomena such as Hawking radiation from black holes and the Unruh effect experienced by accelerated observers in flat spacetime \cite{Hawking:1975vcx,Unruh:1976db}. Hawking radiation emerges from the quantum vacuum near the event horizon of a black hole, leading to thermal emission at a temperature inversely proportional to the black hole mass, while the Unruh effect manifests as a thermal bath perceived by a uniformly accelerated detector in Minkowski vacuum, with temperature proportional to the proper acceleration \cite{Unruh:1976db}. These effects give emphasis to the observer-dependent nature of particle content in quantum fields and have profound implications for black hole thermodynamics and quantum gravity. Experimental verification of such radiation remains elusive due to the extreme conditions required, prompting theoretical explorations of analogous systems and extensions to more complex geometries.

Recent studies have proposed innovative mechanisms to probe acceleration radiation in laboratory-accessible settings, often leveraging analogies between curved spacetime effects and accelerated motion in flat space. The study of radiation from atoms interacting with curved spacetime has forged a deep bridge between quantum optics and black hole thermodynamics. Scully et al. \cite{Scully:2017utk} have shown that atoms freely falling into a black hole radiate in a way reminiscent of laser cooling, offering a quantum–optical lens on Hawking physics. This novel radiation emission, known as Horizon Brightened Acceleration Radiation (HBAR), admits a conformal–quantum–mechanics description and a thermodynamic correspondence with horizon systems \cite{Camblong:2020pme,Azizi:2021qcu,Azizi:2021yto}. It has also been noted that HBAR emission, though bearing analogies with Hawking emission, is endowed with some unique mode structure that sets it distinctly apart from Hawking emission. The framework has broadened to uniformly moving atoms \cite{Svidzinsky:2019jqr}, accelerated atoms in optical cavities \cite{Lopp:2018lxl}, and atom–mirror/entangled setups \cite{Chatterjee:2021fsg,Das:2022qpx}, while dark energy and dark matter backgrounds imprint distinctive nonthermal features on the spectrum \cite{Bukhari:2022wyx,Bukhari:2023yuy}. HBAR entropy analyses for quantum–corrected and charged black holes further test the equivalence principle \cite{Sen:2022tru,Jana:2024fhx}. Recent unifying views recast Unruh, Hawking, and Cherenkov radiation within a common quantum–optical paradigm \cite{Scully:2022pov,Ordonez:2025sqp}. Building on these advances, new results encompass HBAR entropy for atoms infalling into GUP–corrected Schwarzschild backgrounds \cite{Ovgun:2025isv}, a quantum spectral signature at the ISCO \cite{Pantig:2025zhn}, and acceleration radiation from derivative–coupled atoms in modified–gravity black holes \cite{Pantig:2025okn}, collectively underscoring the versatility of quantum–optical tools for horizon physics. For instance, proposals include detecting Unruh-like radiation via electrons in storage rings or ultra-intense laser fields \cite{Bell:1982qr,Chen:1998kp}, as well as simulations using circuit quantum electrodynamics to mimic relativistic trajectories \cite{Johansson:2009zz}. A particularly intriguing approach involves the excitation of a two-level atom undergoing non-uniform acceleration, where virtual vacuum fluctuations are promoted to real photons. In this vein, Dolan, Hunter-McCabe, and Twamley demonstrated that simple harmonic oscillation of a two-level atom in flat Minkowski spacetime, either near a mirror, within a cavity, or in free space, can induce excitation of the atom accompanied by photon emission from the vacuum \cite{Dolan:2020hzm}. Their analysis, based on first-order perturbation theory without the rotating-wave approximation, reveals transition rates that depend on the drive frequency and amplitude, offering a pathway to observable acceleration radiation under non-relativistic conditions. This approach builds on studies of oscillatory detector motion in flat spacetime, extending the analysis to include the nontrivial effects of gravitational curvature \cite{Lin:2017kyr}.

Building upon this framework, we extend the investigation to the curved spacetime of a Schwarzschild black hole, where gravitational curvature introduces additional vacuum structure and potential synergies with acceleration effects. The Schwarzschild metric, an exact solution to Einstein's field equations for a static, spherically symmetric mass, describes the exterior geometry of a non-rotating black hole with an event horizon at $r_{\rm h} = 2M$, where $M$ denotes the black hole's geometric mass \cite{Schwarzschild:1916uq}. We model the atom as an Unruh-DeWitt detector, a point-like two-level system linearly coupled to a massless scalar field, executing small radial oscillations about a fixed mean Schwarzschild radius $R_0>2M$. The worldline is supported (non-geodesic), so an orbital ISCO criterion is irrelevant. What matters for validity is simply that the motion remains nonrelativistic and perturbative around the static worldline at $R_0$, with proper acceleration $a(R_0)=M/\left(R_0^2\sqrt{1-2M/R_0}\right)$, and with the small-amplitude bound $A\Omega\ll \sqrt{f(R_0)}$ where $f(R)=1-2M/R$. Working at fixed $R_0>2M$ also avoids the Boulware pathology at the horizon while keeping the redshift factors explicit in all expressions.
As mentioned, the detector undergoes small-amplitude radial vibrations around $R_0$, with amplitude $A \ll M$, akin to the oscillatory motion in \cite{Dolan:2020hzm} but now along a timelike trajectory in Schwarzschild coordinates. To isolate curvature-induced contributions from Hawking radiation, we employ the Boulware vacuum state for the scalar field, which corresponds to the absence of particles at spatial infinity and is regular at infinity but singular at the horizon \cite{Boulware:1974dm,Birrell:1982ix}. This choice contrasts with the Hartle-Hawking vacuum, which incorporates thermal radiation \cite{Hartle:1976tp}, allowing us to focus on acceleration radiation modulated by the black hole geometry. Previous works have examined Unruh-DeWitt detectors in Schwarzschild spacetime, often for static or orbiting trajectories to probe local temperature or entanglement \cite{Crispino:2007eb,Louko:2007mu,Ng:2014kha,Juarez-Aubry:2014jba, Conroy:2021aow}, but the combination of vibrational motion and Boulware vacuum enables a novel probe of how spacetime curvature amplifies or suppresses vacuum excitation processes akin to those in flat space.

The main new contribution of this work is to provide a fully analytic Floquet treatment of a vibrating Unruh-DeWitt detector in a curved, static black-hole spacetime. We show that the entire harmonic structure of the excitation probability can be written in closed form in terms of Bessel functions with a simple gravitational redshift dressing \(f(R_0)\), so that each Floquet line is shifted and rescaled according to the local gravitational potential. This gives a clean bridge between flat-space vibrating atom proposals \cite{Azizi:2021qcu,Scully:2022bun} and curved-space settings, and allows us to isolate the specific way in which curvature amplifies acceleration radiation. A further novelty is the careful distributional analysis of the finite-time window and of the many-period limit, together with an explicit demonstration that the mode-amplitude picture is exactly equivalent to the standard response-functional/Wightman-function formalism in the present black-hole context.

We organize the paper as follows: In Section \ref{sec2}, we derive the transition probability for detector excitation and scalar particle emission using first-order perturbation theory, adapting the interaction Hamiltonian to the curved background. We obtain closed-form expressions for the emission rate as a function of $R_0$, $A$, and the vibrational frequency, highlighting deviations from the flat-space results due to redshift and near-horizon enhancements. This extension not only bridges acceleration radiation with black hole physics but also suggests potential analogies for condensed-matter simulations of curved spacetimes. We present conclusive remarks in Section \ref{sec6} and state future directions.

\section{Setup and (1+1) Quantization} \label{sec2}
\subsection{Schwarzschild background and (1+1) reduction} \label{sec2.1}
The exterior geometry of a static, spherically symmetric black hole of mass \(M\) is described by the Schwarzschild line element in standard coordinates \((t,r,\theta,\phi)\) with signature \((-,+,+,+)\). In units \(G=c=\hbar=1\), the \begin{equation}
ds^2=-f(r)\,dt^2+f(r)^{-1}\,dr^2+r^2\,(d\theta^2+\sin^2\theta\,d\phi^2),
\end{equation}
where $f(r)=1-\frac{r_h}{r}$ with $r_h\equiv 2M$. When convenient, we fix the length scale by setting \(r_h=1\) (a coordinate rescaling that preserves all local physics). With this choice, \(f(r)=1-1/r\), the event horizon (coordinate singularity) is at \(r=1\), and the asymptotically flat region is \(r\to\infty\).

To analyze radial wave propagation and emphasize the causal structure, it is useful to introduce the tortoise coordinate \(r_*\), defined by
\begin{align}
&\frac{dr_*}{dr}=\frac{1}{f(r)}, \nonumber \\
&r_*(r)=r+r_h\ln\!\left|\frac{r}{r_h}-1\right|
\ \ \stackrel{r_h=1} = \ \
r+\ln|r-1|.
\end{align}
For \(r>r_h\) one has \(r_*\in(-\infty,\infty)\); specifically, \(r_*\to -\infty\) as \(r\rightarrow r_h\) and \(r_*\sim r\) as \(r\to\infty\). In the \((t,r_*)\) chart, the radial part of the metric becomes manifestly conformally flat:
\begin{equation}
ds^2=-f(r)\,\left(dt^2-dr_*^{\,2}\right)+r^2\,d\Omega_2^2,
\end{equation}
where $d\Omega_2^2=d\theta^2+\sin^2\theta\,d\phi^2$. Equivalently, in the null coordinates \(u\equiv t-r_*\) and \(v\equiv t+r_*\), the (1+1) sector is
\begin{equation}
ds^2_{(1+1)}=-f(r)\,du\,dv.
\end{equation}
These forms make explicit that the radial \(t{-}r\) sector is conformally related to (1+1) Minkowski spacetime.

We consider a real, massless scalar field \(\Phi\) on this background. The four‑dimensional action
\begin{equation}
S[\Phi]=-\frac{1}{2}\int d^4x\,\sqrt{-g}\,g^{\mu\nu}\,\partial_\mu\Phi\,\partial_\nu\Phi,
\end{equation}
leads to the Klein–Gordon equation 
\begin{eqnarray}
\square\Phi=\frac{1}{\sqrt{-g}}\partial_\mu\!\left(\sqrt{-g}\,g^{\mu\nu}\partial_\nu\Phi\right)=0,
\end{eqnarray}
for which the separation of variables in spherical harmonics gives
\begin{equation}
\Phi(t,r,\theta,\phi)=\sum_{\ell=0}^{\infty}\sum_{m=-\ell}^{\ell}\frac{\chi_{\ell m}(t,r)}{r}\,Y_{\ell m}(\theta,\phi).
\end{equation}
Each angular sector \((\ell,m)\) obeys a (1+1) wave equation in \((t,r_*)\) with an effective curvature potential of Regge–Wheeler type:
\begin{equation}
\left(-\frac{\partial^2}{\partial t^2}+\frac{\partial^2}{\partial r_*^{\,2}}-V_\ell(r)\right)\chi_{\ell m}(t,r)=0,
\end{equation}
where \(V_\ell(r)\) is a positive barrier that depends on \(\ell\) and \(r\) and vanishes as \(r\to r_h\) and \(r\to\infty\). In the so‑called \(s\)-wave or radial approximation, we retain only the \(\ell=0\) sector and neglect the residual barrier (which is small for suitable frequency windows). This leads to the (1+1) effective field equation
\begin{equation}
\left(-\frac{\partial^2}{\partial t^2}+\frac{\partial^2}{\partial r_*^{\,2}}\right)\chi(t,r)=0,
\end{equation}
for a reduced field \(\chi(t,r)\) defined by \(\Phi\simeq \chi(t,r)/(r\sqrt{4\pi})\) (the \(Y_{00}\) normalization absorbed). In null coordinates \(u,v\), the same equation reads
\begin{equation}
\partial_u\partial_v\,\chi(u,v)=0,
\end{equation}
whose general smooth solution is a superposition \(\chi(u,v)=\chi^{\rm out}(u)+\chi^{\rm in}(v)\) of right‑moving (“outgoing”) and left‑moving (“ingoing”) components.

Quantization in this (1+1) sector proceeds by expanding the Heisenberg field in normal modes of definite Killing frequency \(\omega>0\) with respect to \(\partial_t\). The positive‑frequency plane‑wave solutions are
\begin{align} \label{eq:Boulware_modes}
u^{\rm out}_\omega(t,r)&=\frac{1}{\sqrt{4\pi\omega}}\,e^{-i\omega(t-r_*)}, \nonumber \\
u^{\rm in}_\omega(t,r)&=\frac{1}{\sqrt{4\pi\omega}}\,e^{-i\omega(t+r_*)},
\end{align}
which are positive frequency with respect to \(t\) at spatial infinity and represent, respectively, outward and inward radial propagation in the tortoise coordinate. The full reduced field operator is then
\begin{equation}
\chi(t,r)=\int_0^\infty d\omega\;\left[a^{\rm out}_\omega\,u^{\rm out}_\omega(t,r)
+a^{\rm in}_\omega\,u^{\rm in}_\omega(t,r)
+\text{H.c.}\right],
\end{equation}
with the non‑vanishing commutators \([a^{\rm out}_\omega,a^{\rm out\,\dagger}_{\omega'}]=[a^{\rm in}_\omega,a^{\rm in\,\dagger}_{\omega'}]=\delta(\omega-\omega')\).

The normalization factors \(1/\sqrt{4\pi\omega}\) follow from the Klein–Gordon inner product on a Cauchy slice \(t=\mathrm{const}\). For two solutions \(f,g\) of the (1+1) wave equation,
\begin{equation}
(f,g)=i\int_{-\infty}^{\infty}dr_*\;\left(f^*\,\partial_t g-(\partial_t f^*)\,g\right).
\end{equation}
With this choice, one has \((u^{\rm out}_\omega,u^{\rm out}_{\omega'})=\delta(\omega-\omega')\), \((u^{\rm in}_\omega,u^{\rm in}_{\omega'})=\delta(\omega-\omega')\), and \((u^{\rm out}_\omega,u^{\rm in}_{\omega'})=0\).

A natural vacuum for stationary observers at infinity is the Boulware state, defined by annihilation conditions \(a^{\rm out}_\omega|0\rangle_B=a^{\rm in}_\omega|0\rangle_B=0\) for all \(\omega>0\). Its two‑point (Wightman) function along the radial sector takes the conformally flat (1+1) form
\begin{align}
    G_B^{+}(x,x')&=\langle 0|\,\chi(x)\chi(x')\,|0\rangle_B \nonumber \\
    &=-\frac{1}{4\pi}\ln\!\left[(u-u'-i\epsilon)(v-v'-i\epsilon)\right]+\text{const.},
\end{align}
up to an additive constant that encodes IR regularization in two dimensions. By construction, this Boulware state is empty of particles for static observers at infinity but exhibits divergent stress-energy as \(r\to r_h\) \cite{Birrell:1982ix,Wald:1995yp}. In what follows, we therefore keep the detector at radii \(R_0\) that satisfy the conservative near-horizon validity conditions summarized in Eq. \eqref{eq:validity-domain}, so that the worldline never samples the pathological region. In the present work, we use mode functions and normal ordering with respect to the Boulware basis and evaluate observables along timelike detector worldlines outside the horizon. We emphasize that while the (1+1) reduction captures the causal and redshift structure of radial propagation exactly (through the conformal factor \(f\)), it neglects backscattering and angular‑momentum barriers that appear in the full (3+1) theory; these enter as density‑of‑states and greybody factors in a higher‑dimensional treatment and are briefly discussed in Section \ref{sec5}.

\subsection{Detector model, trajectory, and switching} \label{sec2.2}
We model the atom as a two‑level Unruh–DeWitt (UDW) detector linearly coupled to a real, massless scalar field along a prescribed timelike worldline. The detector's Hilbert space is spanned by an orthonormal pair of internal states, denoted \(|g\rangle\) (ground) and \(|e\rangle\) (excited), with energy gap \(\omega_0>0\) defined in the detector's proper time. The free detector Hamiltonian in the interaction picture is encoded through the monopole operator
\begin{equation}
\mu(\tau)=\sigma_{+}\,e^{i\omega_0\tau}+\sigma_{-}\,e^{-i\omega_0\tau},
\end{equation}
where \(\sigma_{+}=|e\rangle\langle g|\) and \(\sigma_{-}=|g\rangle\langle e|\), so that \(\langle e|\sigma_{+}|g\rangle=1=\langle g|\sigma_{-}|e\rangle\), with all other matrix elements vanishing. We work without the rotating‑wave approximation (RWA) to retain counter‑rotating terms that are essential for acceleration radiation.

The detector couples to the field through a scalar monopole interaction localized on its worldline \(x^\mu(\tau)\). Introducing a real switching function \(\chi(\tau)\) that modulates the interaction in time, the interaction Hamiltonian density reads
\begin{equation} \label{eq:int_hamiltonian}
H_I(\tau)=g\,\chi(\tau)\,\mu(\tau)\,\Phi\left(x(\tau)\right),
\end{equation}
with \(g\) a dimensionless coupling constant. Throughout this work, \(\Phi\) denotes the reduced (1+1) field obtained from the \(s\)-wave sector as explained in Section \ref{sec2.1}. We work in units $G=c=\hbar=1$. In $(1{+}1)$ dimensions with Klein–Gordon normalization $u_\omega=(4\pi\omega)^{-1/2}e^{-i\omega(t\mp r_*)}$, the detector–field coupling $g$ is dimensionless in our conventions. Transition probabilities reported below are per driving period (or, where stated, rates) and are therefore dimensionless. Prefactors such as $1/\omega_n$ arise from the KG normalization and the Jacobian when imposing the resonance condition and do not indicate a dimensional mismatch \cite{DeWitt:1979,Crispino:2008}.

The detector's worldline is taken to be radial, confined to the equatorial plane \(\theta=\pi/2\), \(\phi=0\), and parametrized by proper time \(\tau\). We consider small‑amplitude harmonic vibrations about a fixed mean radius \(R_0\) outside the horizon, with \(R_0 > r_{\rm ISCO}\) to avoid circular‑orbit instabilities. The radial coordinate is prescribed as
\begin{equation}\label{trajectory}
r(\tau)=R_0+A\cos(\Omega\tau),
\end{equation}
where \(A\ll R_0\) sets the oscillation amplitude and \(\Omega>0\) is the angular frequency measured in proper time. The detector's four‑velocity satisfies \(u^\mu u_\mu=-1\). Using the Schwarzschild line element, the relation between coordinate time \(t\) and proper time \(\tau\) is
\begin{align}
-1&=-f(r)\,\dot{t}^{\,2}+f(r)^{-1}\,\dot{r}^{\,2} \nonumber \\
&\Rightarrow
\frac{dt}{d\tau}=\frac{1}{\sqrt{f(r)}}\sqrt{1+\frac{\dot r^{\,2}}{f(r)}},
\end{align}
where dots denote \(d/d\tau\).  The radial speed must obey \(|\dot r|<\sqrt{f(r)}\). With the trajectory defined in Eq.(\ref{trajectory}), the condition 
\begin{equation}\label{condition}
|\dot{r}|\sim A\Omega\ll \sqrt{f(R_0)},
\end{equation}
suffices to ensure small‑amplitude, adiabatic‑velocity regime. With that, it follows
\begin{equation}
\frac{dt}{d\tau}\simeq f(R_0)^{-1/2}+ \mathcal{O}(A^2),\quad
t(\tau)\simeq t_0+\frac{\tau}{\sqrt{f(R_0)}}+\mathcal{O}(A^2),
\end{equation}
with an irrelevant constant \(t_0\). Introducing the tortoise coordinate \(r_*\) defined by \(dr_*/dr=f(r)^{-1}\), we expand to leading order in \(A\) as
\begin{align}
r_*(\tau)&=r_*\left(R_0\right)+\int_{R_0}^{r(\tau)}\frac{dr}{f(r)} \nonumber \\
& \simeq r_*\left(R_0\right)+\frac{A}{f(R_0)}\cos(\Omega\tau)+\mathcal{O}(A^2).
\end{align}
It will be convenient to record the corresponding null coordinates along the worldline, \(u(\tau)=t(\tau)-r_*(\tau)\) and \(v(\tau)=t(\tau)+r_*(\tau)\), whose proper‑time derivatives are
\begin{align}
\frac{du}{d\tau}&\simeq \frac{1}{\sqrt{f(R_0)}}+\frac{A\Omega}{f(R_0)}\sin(\Omega\tau),\\
\frac{dv}{d\tau}&\simeq \frac{1}{\sqrt{f(R_0)}}-\frac{A\Omega}{f(R_0)}\sin(\Omega\tau),
\end{align}
again up to \(\mathcal{O}(A^2)\). These expressions will enter the phase evolution of positive‑frequency modes evaluated on the detector's worldline. To maintain timelike motion at leading order in \(A\), our system must satisfy the condition in Eq.\ref{condition}. In what follows, we consistently truncate at \(\mathcal{O}(A)\) in kinematical expansions and at leading order in \(g\) in the perturbative transition probabilities.

We now specify the switching function \(\chi(\tau)\) used to define the detector's response. Two equivalent and standard choices are convenient: First is the finite rectangular support over one oscillation period \(T=2\pi/\Omega\),
\begin{equation}
\chi_T(\tau)=\Theta(\tau)\,\Theta(T-\tau),
\end{equation}
which isolates a single cycle of the periodic trajectory. The accumulated probability over that cycle, divided by \(T\), defines a period‑averaged quantity. Second is the smooth long‑time periodic switching, e.g.,
\begin{equation}
    \chi(\tau)=\sum_{k=-N}^{N}s\left((\tau-kT)/\Delta\right),
\end{equation}
with a bump function \(s\) of width \(\Delta\ll T\) and \(N\gg1\), followed by the limit \(N\to\infty\) at fixed \(T\). In this case, one averages over a large number of identical periods and divides by the total observation time \(T_{\rm obs}\approx (2N+1)T\).

Both procedures lead to the same Floquet‑type resonance structure: the periodic motion imprints a discrete frequency comb with harmonics at integer multiples of \(\Omega\), and the per‑period average isolates the corresponding coefficients. In this work, we adopt the first choice for transparency. Concretely, if \(\mathcal{A}_T(\omega)\) denotes the first‑order transition amplitude into a field mode of Killing frequency \(\omega\) accumulated over a single period, the probability per period is \(|\mathcal{A}_T(\omega)|^2\). The period‑averaged quantity that we report in Section \ref{sec3} is then defined by dividing by \(T\),
\begin{equation}
\overline{P}(\omega)\equiv \frac{|\mathcal{A}_T(\omega)|^2}{T}.
\end{equation}
When the resonance condition selects discrete frequencies \(\omega=\omega_n\) labeled by an integer \(n\ge1\), we write \(\overline{P}_n\equiv \overline{P}(\omega_n)\). This convention matches the usage in the flat‑space vibrating‑detector case \cite{Dolan:2020hzm}, and is convenient for comparing curved‑space redshift effects against the Minkowski limit.

\subsection{Boulware modes and normalization} \label{sec2.3}
The Schwarzschild exterior admits a static Killing field \(\partial_t\), and the natural “empty” state for static observers at spatial infinity is the Boulware vacuum \cite{Boulware:1974dm,Birrell:1982ix}. It is defined by declaring positive frequency mode solutions to those which are are positive with respect to the Killing time \(t\) at \(\mathcal{I}^-\) and \(\mathcal{I}^+\), and by annihilating the corresponding annihilation operators on the vacuum. In the (1+1) radial reduction (Section \ref{sec2.1}), the reduced field \(\chi\) decomposes into two independent sectors: outgoing (“right‑moving”) and ingoing (“left‑moving”) waves along the tortoise coordinate \(r_*\). The associated positive‑frequency $(\omega > 0)$ Boulware modes are plane waves in the null coordinates \(u=t-r_*\) and \(v=t+r_*\) is given by Eq. \eqref{eq:Boulware_modes}. The overall factor \(1/\sqrt{4\pi\omega}\) is the normalization constant chosen so that these modes form a \(\delta\)-orthonormal basis with respect to the Klein–Gordon (KG) inner product on a constant‑\(t\) Cauchy slice. Explicitly, for any two on‑shell solutions \(f,g\) of the (1+1) wave equation,
\begin{equation}
(f,g)\;=\;i\int_{-\infty}^{\infty}dr_*\;\left(f^*\,\partial_t g-(\partial_t f^*)\,g\right),
\end{equation}
and the normalization requirements read
\begin{align}
\left(u^{\rm out}_\omega,u^{\rm out}_{\omega'}\right)&=\delta(\omega-\omega')\,
\left(u^{\rm in }_\omega,u^{\rm in }_{\omega'}\right) \nonumber \\
&=\delta(\omega-\omega')\,
\left(u^{\rm out}_\omega,u^{\rm in }_{\omega'}\right)=0,
\end{align}
together with the negative‑norm relations for the complex‑conjugate modes,
\begin{align}
\left(u^{\rm out\,*}_\omega,u^{\rm out\,*}_{\omega'}\right)&=-\delta(\omega-\omega'),\nonumber \\
\left(u^{\rm in \,*}_\omega,u^{\rm in \,*}_{\omega'}\right)&=-\delta(\omega-\omega').
\end{align}

It is instructive to verify the normalization constant. Consider the outgoing sector; at fixed \(t\) one has
\begin{equation}
u^{\rm out}_\omega(t,r)=N\,e^{-i\omega t}\,e^{+i\omega r_*},\quad
\partial_t u^{\rm out}_{\omega'}=-i\omega'\,u^{\rm out}_{\omega'}.
\end{equation}
Then
\begin{align}\nonumber
\left(u^{\rm out}_\omega,u^{\rm out}_{\omega'}\right)
&=i\!\int\!dr_*\;\left[u^{\rm out *}_\omega(-i\omega')u^{\rm out}_{\omega'}
-(+i\omega)\,u^{\rm out *}_\omega u^{\rm out}_{\omega'}\right]\\
&=(\omega+\omega')|N|^2\!\int\!dr_*\;e^{i(\omega-\omega')r_*}.
\end{align}
Using \( \int_{-\infty}^{\infty}dr_*\;e^{i(\omega-\omega')r_*}=2\pi\,\delta(\omega-\omega'), \) and enforcing \(\left(u^{\rm out}_\omega,u^{\rm out}_{\omega'}\right)=\delta(\omega-\omega')\) gives \( (\omega+\omega')|N|^2\,2\pi\,\delta(\omega-\omega')=\delta(\omega-\omega'). \) On the support of the delta function one has \(\omega'=\omega\), so \((\omega+\omega')\to 2\omega\), and therefore
\begin{equation}
|N|^2=\frac{1}{4\pi\,\omega}\quad\Rightarrow\quad N=\frac{1}{\sqrt{4\pi\,\omega}}.
\end{equation}
The same computation holds for the ingoing sector, with \(e^{-i\omega(t+r_*)}\) replacing \(e^{-i\omega(t-r_*)}\). The vanishing cross product \(\left(u^{\rm out}_\omega,u^{\rm in }_{\omega'}\right)=0\) follows from an integral of \(e^{i(\omega+\omega')r_*}\), which yields \(\delta(\omega+\omega')=0\) for \(\omega,\omega'>0\), hence orthogonality of the sectors.

With these conventions, the reduced field and its canonical conjugate admit the mode expansion
\begin{equation}
    \chi(t,r)=\int_0^\infty d\omega\;\left[a^{\rm out}_\omega\,u^{\rm out}_\omega(t,r) +a^{\rm in }_\omega\,u^{\rm in }_\omega(t,r)+\text{H.c.}\right],
\end{equation}
and the canonical commutation relations are implemented by
\begin{equation}
    \left[a^{\rm out}_\omega,a^{\rm out\,\dagger}_{\omega'}\right]=\delta(\omega-\omega'),\quad \left[a^{\rm in }_\omega,a^{\rm in \,\dagger}_{\omega'}\right]=\delta(\omega-\omega'),
\end{equation}
with all other commutators vanishing. The Boulware vacuum is characterized by
\begin{equation}
    a^{\rm out}_\omega\,|0\rangle_B=0,\quad a^{\rm in }_\omega\,|0\rangle_B=0,\quad \forall\,\omega>0.
\end{equation}
Equivalently, \(|0\rangle_B\) is the Fock vacuum associated with positive Killing‑frequency modes at infinity in both the outgoing and ingoing sectors. This state is regular at spatial infinity but is singular on the horizon in (3+1) (its renormalized stress tensor diverges there); in the present work, we evaluate detector observables on worldlines at radii \(R_0>r_h\), away from the horizon, where mode normal ordering in the Boulware basis is well defined \cite{Kay:2015iwa}.

For later use, it is convenient to write the Boulware Wightman function in the (1+1) reduction as a mode integral. Along generic points \(x=(t,r)\), \(x'=(t',r')\),
\begin{align}
    G_B^{+}(x,x')&=\langle 0|\chi(x)\chi(x')|0\rangle_B \nonumber \\ 
    &=\int_0^\infty\frac{d\omega}{4\pi\,\omega}\Big[e^{-i\omega\left((t-t')-(r_*-r_*')-i\epsilon\right)} \nonumber \\
    &+e^{-i\omega\left((t-t')+(r_*-r_*')-i\epsilon\right)}\Big].
\end{align}
In null coordinates \(u=t-r_*\), \(v=t+r_*\), this becomes
\begin{equation}
    G_B^{+}(x,x')=\int_0^\infty\frac{d\omega}{4\pi\,\omega}\left[e^{-i\omega\left((u-u')-i\epsilon\right)}+e^{-i\omega\left((v-v')-i\epsilon\right)}\right].
\end{equation}
The integrals can be evaluated in the distributional sense using the identities \(\int_0^\infty d\omega\,e^{-i\omega s}/\omega= -\ln(\mu|s|)-i\frac{\pi}{2}\,\mathrm{sgn}\,s\) up to an arbitrary IR scale \(\mu\). This yields the familiar conformal form
\begin{equation}
    G_B^{+}(x,x')=-\frac{1}{4\pi}\,\ln\!\left[(u-u'-i\epsilon)(v-v'-i\epsilon)\right]+\text{const.},
\end{equation}
up to an additive constant that reflects the infrared regularization intrinsic to massless fields in two dimensions. What matters for transition probabilities is the \((u,v)\)-dependent part; additive constants drop out of physically regulated detector observables.

We now discuss the meaning of the factor \(1/\sqrt{4\pi\omega}\). First, it enforces KG \(\delta\)-normalization of the stationary modes so that the creation and annihilation operators satisfy the standard continuous‑spectrum algebra, which in turn guarantees the canonical equal‑time commutators of the field. Second, it is equivalent to unit‑flux normalization of the plane waves in the asymptotic regions when interpreted via the conserved KG current; in particular, the outgoing mode \(u^{\rm out}_\omega\) carries a unit delta‑normalized flux toward \(\mathcal{I}^+\) and the ingoing mode \(u^{\rm in}_\omega\) carries a unit delta‑normalized flux inward. These properties underlie the mode‑by‑mode transition amplitudes used in Section \ref{sec3} and make explicit how the frequency measure \(d\omega\) combines with the normalization to produce the characteristic factor \(1/(4\pi\omega)\) that appears in the period‑averaged probability \(\overline{P}_n^{(1+1)}\). References for these standard constructions in curved‑spacetime QFT include classic treatments of Schwarzschild mode normalization and Boulware quantization, as well as general discussions of the KG inner product and Wightman functions.

\section{Transition Probability in (1+1)} \label{sec3}
\subsection{Mode-amplitude derivation} \label{sec3.1}
We work at first order in the coupling, in the interaction picture, and consider the transition \(|g,0\rangle \to |e,1_\omega\rangle\) with the detector moving on the oscillatory worldline specified in Section \ref{sec2.2} and the field quantized in Boulware modes of Section \ref{sec2.3}. The interaction Hamiltonian along the worldline is given by Eq. \eqref{eq:int_hamiltonian}. The relevant matrix element comes from the \(\sigma_{+}e^{i\omega_0\tau}\) part of \(\mu(\tau)\) and the creation part of \(\Phi\). With a finite switching window over one period \(T\equiv 2\pi/\Omega\) (rectangular switching \(\chi_T(\tau)=\Theta(\tau)\Theta(T-\tau)\)), the first-order amplitude into a mode of Killing frequency \(\omega>0\) is
\begin{align}\nonumber
    \mathcal{A}_T(\omega)&= g\int_0^{T} d\tau\, e^{\,i\omega_0\tau}\,\langle 1_\omega|\Phi(x(\tau))|0\rangle\\
   \label{AMPLITUDE_1} &= g\int_0^{T} d\tau\, e^{\,i\omega_0\tau}\,u_\omega^{*}(x(\tau)).
\end{align}
We focus on the outgoing sector, for which
\begin{align}
    u_\omega^{\rm out}(x(\tau))&=\frac{1}{\sqrt{4\pi\omega}}\,\exp\!\big\{\,-i\omega\,[t(\tau)-r_*(\tau)]\big\},\\ 
    u_\omega^{\rm out\,*}(x(\tau))&=\frac{1}{\sqrt{4\pi\omega}}\,\exp\!\big\{\,i\omega\,[t(\tau)-r_*(\tau)]\big\}.
\end{align}
For completeness, the ingoing sector is obtained by replacing $u=t-r_*$ with $v=t+r_*$.
To leading order in the small-amplitude expansion around $R_0$ we have
\begin{align}
t(\tau)&\simeq t_0+\frac{\tau}{\sqrt{f(R_0)}},\nonumber \\
r_*(\tau)&\simeq r_*(R_0)+\frac{A}{f(R_0)}\cos(\Omega\tau),
\end{align}
so
\begin{equation}
v(\tau)=t(\tau)+r_*(\tau)\simeq \mathrm{const}
+ \frac{\tau}{\sqrt{f(R_0)}}
+ \frac{A}{f(R_0)}\cos(\Omega\tau).
\end{equation}
Inserting $v(\tau)$ in the ingoing Boulware modes yields the same resonance condition
$\alpha(\omega)+m\Omega=0$ and the same Bessel weights $J_m(\cdot)$ (phases differ but do not affect per-period probabilities), hence all formulas below carry over to the ingoing sector unchanged. The ingoing Boulware sector proceeds identically and produces the same resonance set and Bessel weights, differing only by overall phases that do not affect per-period probabilities. Using the kinematical expansions from Section \ref{sec2.2} to \(\mathcal{O}(A)\),
\begin{equation}
    t(\tau)\simeq t_0+\frac{\tau}{\sqrt{f(R_0)}},\quad r_*(\tau)\simeq r_*(R_0)+\frac{A}{f(R_0)}\cos(\Omega\tau),
\end{equation}
the amplitude (\ref{AMPLITUDE_1}) becomes
\begin{equation}
    \mathcal{A}_T(\omega)=\frac{g}{\sqrt{4\pi\omega}}\,e^{\,i\varphi_0}\int_0^{T}\! d\tau\;e^{\,i\alpha(\omega)\,\tau - i z(\omega)\,\cos(\Omega\tau)},
\end{equation}
where overall constants have been collected into an irrelevant phase \(e^{i\varphi_0}\), and we have defined
\begin{equation}
    \alpha(\omega)\equiv \omega_0+\frac{\omega}{\sqrt{f(R_0)}},\quad z(\omega)\equiv \frac{\omega A}{f(R_0)}.
\end{equation}

To proceed, we use the Jacobi–Anger identity for a sinusoidal phase modulation,
\begin{equation}
    e^{-i\,z\cos(\Omega\tau)}=\sum_{m=-\infty}^{\infty}(-i)^m\,J_m(z)\,e^{\,i m\Omega\tau},
\end{equation}
with \(J_m\) the Bessel function of the first kind. Substituting into the integral yields
\begin{align}
    \mathcal{A}_T(\omega)&=\frac{g}{\sqrt{4\pi\omega}}\,e^{\,i\varphi_0}\sum_{m=-\infty}^{\infty}(-i)^m J_m\!\left(z(\omega)\right) \nonumber \\
    &\times \,\underbrace{\int_0^{T}\! d\tau\; e^{\,i[\alpha(\omega)+m\Omega]\tau}}_{\displaystyle I_T(\alpha(\omega)+m\Omega)}.
\end{align}
The time integral is elementary,
\begin{equation}
    I_T(\beta)=\int_0^{T} d\tau\,e^{i\beta\tau}=e^{\,i\beta T/2}\,T\,\mathrm{sinc}\!\left(\frac{\beta T}{2}\right), 
\end{equation}
where we note $\mathrm{sinc}(x)\equiv \frac{\sin x}{x}$.  Therefore,
\begin{align}
    \mathcal{A}_T(\omega)&=\frac{g}{\sqrt{4\pi\omega}}\,e^{\,i\varphi_0}\sum_{m=-\infty}^{\infty}i^m J_m\!\left(z(\omega)\right)\,e^{\,i[\alpha(\omega)+m\Omega]T/2}\,T\, \nonumber \\
    & \times \mathrm{sinc}\!\left(\frac{[\alpha(\omega)+m\Omega]T}{2}\right).
\end{align}

At this stage, two equivalent viewpoints are useful. First, for a single finite period \(T=2\pi/\Omega\), the sinc kernel is sharply peaked whenever \(\alpha(\omega)+m\Omega\) is near zero. The dominant contributions are near the discrete “Floquet” values of \(\omega\) that satisfy \(\alpha(\omega)+m\Omega=0\). We will exploit this shortly to read off the resonances and their weights. Second, to make the comb of resonances explicit in the distributional sense, consider a long observation time \(T_{\rm obs}\) that covers many periods (or a smooth long-time periodic switching). In the limit \(T_{\rm obs}\to\infty\) one has
\begin{equation}
    \lim_{T_{\rm obs}\to\infty} e^{\,i\beta T_{\rm obs}/2}\,T_{\rm obs}\,\mathrm{sinc}\!\left(\frac{\beta T_{\rm obs}}{2}\right)=2\pi\,\delta(\beta),
\end{equation}
so the sum over \(m\) is promoted to a Dirac-comb of delta functions, enforcing the resonance conditions
\begin{equation}
    \alpha(\omega)+m\Omega=0 \Longleftrightarrow \omega=\omega_m,
\end{equation}
where $\omega_m \equiv -\,m\,\Omega\,\sqrt{f(R_0)}-\omega_0\,\sqrt{f(R_0)}$. Since we require \(\omega>0\), only terms with \(m=-n\) and \(n\in\mathbb{N}\) satisfying \(n\Omega>\omega_0\) contribute. Writing \(m=-n\), the resonant frequencies are
\begin{equation}
    \omega_n=\sqrt{f(R_0)}\,(n\Omega-\omega_0),\quad n=1,2,\dots,\quad n\Omega>\omega_0.
\end{equation}
At each such resonance, we can evaluate the Bessel coefficient at 
\begin{equation}
    z(\omega_n) = \frac{\omega_n A}{f(R_0)} = \frac{(n\Omega-\omega_0)A}{\sqrt{f(R_0)}}.
\end{equation}
The coefficient multiplying the resonance \(m=-n\) is
\begin{equation}
    (-1)^{-n}J_{-n}\!\left(z(\omega_n)\right)=J_n\!\left(z(\omega_n)\right),
\end{equation}
where we used \(J_{-n}(z)=(-1)^n J_n(z)\). Thus, in the long-time limit, the amplitude selects the discrete set \(\{\omega_n\}\) with weights proportional to \(J_n\left(z(\omega_n)\right)\).

Returning to the single-period window \(T=2\pi/\Omega\), we can evaluate the on-resonance contribution by setting \(\alpha(\omega_n)-n\Omega=0\) in the phase and sinc kernel. The time integral reduces to \(I_T(0)=T\), and the amplitude contribution from the \(m=-n\) harmonic is
\begin{equation}
    \mathcal{A}_T(\omega_n)=\frac{g}{\sqrt{4\pi\omega_n}}\,e^{\,i\varphi_n}\,J_n\!\left(\frac{(n\Omega-\omega_0)A}{\sqrt{f(R_0)}}\right)\,T,
\end{equation}
where \(e^{i\varphi_n}\) collects a global phase that is irrelevant for probabilities. Off resonance, the contribution of each \(m\) is suppressed by the sinc factor \(\mathrm{sinc}([\alpha(\omega)+m\Omega]T/2)\), which becomes increasingly narrow as the number of oscillation periods increases.

Two remarks are in order. First, if one includes the ingoing sector \(u^{\rm in}_\omega\), the algebra is identical with \(t(\tau)+r_*(\tau)\) in place of \(t(\tau)-r_*(\tau)\). The same resonance condition \(\alpha(\omega)+m\Omega=0\) follows, with analogous Bessel weights; in this paper, we keep the outgoing sector to model quanta that reach \(\mathcal{I}^+\) in a (3+1) completion. Second, for small argument \(z\ll 1\), the harmonic weights obey \(J_n(z)\sim (z/2)^n/n!\), so higher harmonics are parametrically suppressed in the regime \(A\ll 1\) and moderate \(n\), while the redshift factor \(1/\sqrt{f(R_0)}\) enhances the effective argument \(z(\omega_n)\) as \(R_0\) decreases.

\subsection{From finite-time window to the Floquet transition rate expression} \label{sec3.2}
This subsection turns the finite-time transition amplitude obtained in Section \ref{sec3.1} into a clean, Floquet transition rate. The key steps are: (i) keep the single-period rectangular window of duration $T\equiv 2\pi/\Omega$; (ii) evaluate the on-resonance contribution where the sinc kernel is maximal; (iii) divide by the duration $T$ to define the per-period average. This procedure isolates the Floquet harmonics and makes the factor $2\pi/\Omega$ appear transparently.

Recall the single-period amplitude into an outgoing Boulware mode of Killing frequency $\omega>0$ (Section \ref{sec3.1}):
\begin{align} \label{eq:sing_per_A}
A_T(\omega)
&=\frac{g}{\sqrt{4\pi\omega}}\,
e^{i\phi_0}\sum_{m\in\mathbb{Z}}(-i)^mJ_m\!\left(z(\omega)\right)\;
e^{\frac{i}{2}[\alpha(\omega)+m\Omega]T}\;
T\, \nonumber \\
 & \quad \times \mathrm{sinc}\!\left(\frac{[\alpha(\omega)+m\Omega]T}{2}\right),
\end{align}
where
\begin{equation}
\alpha(\omega)=\omega_0+\frac{\omega}{\sqrt{f(R_0)}},\quad
z(\omega)=\frac{\omega A}{f(R_0)}.
\end{equation}

The finite-time kernel $T\,\mathrm{sinc}([\alpha(\omega)+m\Omega]T/2)$ is sharply peaked whenever \( \alpha(\omega)+m\Omega=0, \) with width $\Delta\beta\sim 2\pi/T$ in the variable $\beta=\alpha(\omega)+m\Omega$. In the many-period limit, this kernel tends in the distributional sense to \(T\,\mathrm{sinc}([\alpha(\omega)+m\Omega]T/2)\to 2\pi\,\delta(\alpha(\omega)+m\Omega)\), as justified in Appendix \ref{appendixA}, thereby producing a Dirac-comb of Floquet resonances. The corresponding Floquet resonances \cite{magnus2013hill} occur at
\begin{equation}
    \omega=\omega_m=-\sqrt{f(R_0)}\,[\,\omega_0+m\Omega\,].
\end{equation}
Because $\omega>0$, only $m=-n$ with $n\in\mathbb{N}$ and $n\Omega>\omega_0$ contribute, giving
\begin{equation} \label{eq:killing_freq}
    \omega_n=\sqrt{f(R_0)}\,(n\Omega-\omega_0), \quad n=1,2,\dots,\ \ n\Omega>\omega_0.
\end{equation}
In the response-functional derivation of Appendix \ref{appendixB}, the same set of resonant frequencies arises from the zeros of the combination \(\omega_0-\omega/\sqrt{f(R_0)}+m\Omega\); the equivalence with the condition \(\alpha(\omega)+m\Omega=0\) used here follows from the reparametrization spelled out in Eq. \eqref{eq:A14}. Near $\omega=\omega_n$, only the harmonic $m=-n$ survives; using $J_{-n}=(-1)^nJ_n$, the on-resonance amplitude over a single period reduces to
\begin{equation} \label{eq:on-reso}
    A_T(\omega_n)=\frac{g}{\sqrt{4\pi\omega_n}}\,e^{i\phi_n}\,J_n\!\left(\frac{(n\Omega-\omega_0)A}{\sqrt{f(R_0)}}\right)\,T.
\end{equation}
Thus $|A_T(\omega_n)|^2\propto T^2$. The growth $\propto T$ in probability (after integrating over a narrow bandwidth) or $\propto T^2$ in $|A_T|^2$ is the usual signature of phase-coherent driving at resonance.

For a finite observation window, a natural way to remove the trivial linear growth with $T$ is to average over that window. We therefore define the per-period (i.e., period-averaged) quantity introduced in Section \ref{sec2.2},
\begin{equation}
    \overline{P}(\omega)\equiv \frac{|A_T(\omega)|^2}{T}.
\end{equation}
Evaluated at a resonant $\omega=\omega_n$, Eq. \eqref{eq:on-reso} gives directly
\begin{equation}
    \overline{P}(\omega_n)
=\frac{g^2}{4\pi\omega_n}
\left[J_n\!\left(\frac{(n\Omega-\omega_0)A}{\sqrt{f(R_0)}}\right)\right]^2
\;T.
\end{equation}
Finally, inserting the single-period duration
\begin{equation}
    T=\frac{2\pi}{\Omega}
\end{equation}
yields the factor as
\begin{align} \label{eq:P_n}
    \overline{P}^{(1+1)}_n
    &\equiv P(\omega_n)
    =\frac{2\pi}{\Omega}\;\frac{g^2}{4\pi\,\omega_n}\;
    J_n^2\!\left(\frac{(n\Omega-\omega_0)A}{\sqrt{f(R_0)}}\right)\, \nonumber \\
    &\times \Theta(n\Omega-\omega_0),\
\end{align}
where $\omega_n=\sqrt{f(R_0)}(n\Omega-\omega_0).$
Here $\Theta$ is the Heaviside step function, making explicit that only harmonics with $n\Omega>\omega_0$ are open. The period-average replaces the explicit $T$ proportionality by $2\pi/\Omega$. No further limiting procedure is required; the result follows from evaluating the resonant term over a single cycle and dividing by its duration. From here, we call Eq. \eqref{eq:P_n} the Floquet transition rate.

The derivation of Eq. \eqref{eq:P_n} also makes clear the role of normalization, where the prefactor $1/\sqrt{4\pi\omega_n}$ originates from the Boulware KG normalization of modes (Section \ref{sec2.3}). The only dependence on the switching window that survives the average is the duration $T$; finer details (rectangular vs smooth) do not affect the per-period result at resonance (see Appendix \ref{appendixA} for a distributional proof).

For completeness, one can recover the same factor via the long-time limit and $\delta$-normalization. For this recovery, we replace the single-period window by $N\gg1$ identical periods (or a smooth periodic switching) with total duration $T_{\text{obs}}=(2N+1)T$. In this limit,
\begin{equation}
    \lim_{T_{\text{obs}}\to\infty} e^{i\beta T_{\text{obs}}/2}T_{\text{obs}}\;\mathrm{sinc}\!\left(\frac{\beta T_{\text{obs}}}{2}\right)=2\pi\delta(\beta),
\end{equation}
so the $m$-sum becomes a Dirac comb $\sum_m J_m(\cdots)\,\delta(\alpha+m\Omega)$ that enforces the same resonances $\omega=\omega_n$.

Next, the integrated probability accumulated over $T_{\text{obs}}$ is then
\begin{align}\nonumber
P_{\text{obs}}&=\int_0^\infty\!d\omega\,|A_{T_{\text{obs}}}(\omega)|^2\\
&=\sum_{n\ge1}\frac{g^2}{4\pi\,\omega_n}\,J_n^2\!\left(\frac{(n\Omega-\omega_0)A}{\sqrt{f(R_0)}}\right)\;2\pi\;T_{\text{obs}}\sqrt{f(R_0)}.
\end{align}
Here we used $\delta(\alpha+m\Omega)=\delta\!\left(\omega/\sqrt{f}+(\omega_0+m\Omega)\right)=\sqrt{f}\,\delta(\omega-\omega_m)$ and evaluated $1/\omega$ at $\omega_n$. Dividing by observation time gives the time-averaged rate $P_{\text{obs}}/T_{\text{obs}}$, which matches Eq. \eqref{eq:P_n} once one identifies a per-period average over a single cycle by substituting $T=2\pi/\Omega$ and restricting to one period.

Either route, single-period averaging (direct and simple) or long-time $\delta$-comb normalization (distributionally precise), produces the same, universal prefactor $2\pi/\Omega$ in the period-averaged expression.

Finally, when the observation is restricted to a single cycle, the sinc kernel in Eq. \eqref{eq:sing_per_A} has finite width $\Delta\omega\sim 2\pi/(T\sqrt{f})=\Omega/\sqrt{f}$. Thus, away from exact resonance,
\begin{equation}
    P(\omega)\sim \frac{g^2}{4\pi\omega T}\left[\sum_m J_m\!\left(z(\omega)\right)\;T\,\mathrm{sinc}\!\left(\frac{[\alpha(\omega)+m\Omega]T}{2}\right)\right]^2
\end{equation}
is suppressed by $\mathrm{sinc}^2$, and the area under each peak is independent of the detailed shape of the window in the large-$Q$ (weak damping) sense. As $N$ periods are concatenated, peaks narrow as $1/N$ while their heights grow as $N^2$, keeping the per-period area fixed and reproducing Eq. \eqref{eq:P_n} in the limit.

\subsection{Final closed form \texorpdfstring{$\overline{P}^{(1+1)}_n$}{}} \label{sec3.3}
In this subsection, we collect the ingredients of Sections \ref{sec3.1}-\ref{sec3.2} and present the closed-form, Floquet transition rate for the $n$-th Floquet harmonic of the detector's sinusoidal motion, together with the precise assumptions under which the expression holds.

For a two–level Unruh–DeWitt detector with energy gap $\omega_0>0$, linearly coupled (with coupling $g$) to a massless scalar field in the Schwarzschild Boulware vacuum, undergoing small-amplitude radial oscillations about a fixed mean radius $R_0$,
\begin{equation}
    r(\tau)=R_0+A\cos(\Omega\tau),\quad 0<A\ll R_0,\quad \Omega>0,
\end{equation}
the Floquet transition rate (per oscillation period) into an outgoing Boulware mode resonant with the $n$-th Floquet line is
\begin{equation} \label{eq:per_harmonic}
	\overline{P}^{(1+1)}_n
	=\frac{2\pi}{\Omega}\,\frac{g^2}{4\pi\,\omega_n}\;
	J_n^2\!\!\left(\frac{(n\Omega-\omega_0)\,A}{\sqrt{f(R_0)}}\right),
\end{equation}
where $\omega_n=\sqrt{f(R_0)}\,(n\Omega-\omega_0),\quad n\in\mathbb{N},\ n\Omega>\omega_0.$
Equivalently, this is
\begin{equation} \label{eq:per-period}
    \overline{P}^{(1+1)}_n
=\frac{g^2}{2\,\Omega\,\omega_n}\;
J_n^2\!\!\left(\frac{(n\Omega-\omega_0)\,A}{\sqrt{f(R_0)}}\right),
\end{equation}
where $f(R)\equiv 1-\frac{2M}{R}$ and $J_n$ is the Bessel function of the first kind. A Heaviside selection rule is implicit:
\begin{equation}
    \overline{P}^{(1+1)}_n\propto \Theta(n\Omega-\omega_0),
\end{equation}
i.e., only harmonics with $n\Omega>\omega_0$ are kinematically allowed (otherwise $\omega_n<0$ and no outgoing mode exists at that frequency). The total Floquet transition rate into the outgoing sector is therefore
\begin{equation}\label{eq:totalFR}
    \overline{P}^{(1+1)}_{\rm out}=\sum_{n=1}^{\infty}\overline{P}^{(1+1)}_n\;\Theta(n\Omega-\omega_0).
\end{equation}
\begin{figure}
    \centering
    \includegraphics[width=\columnwidth]{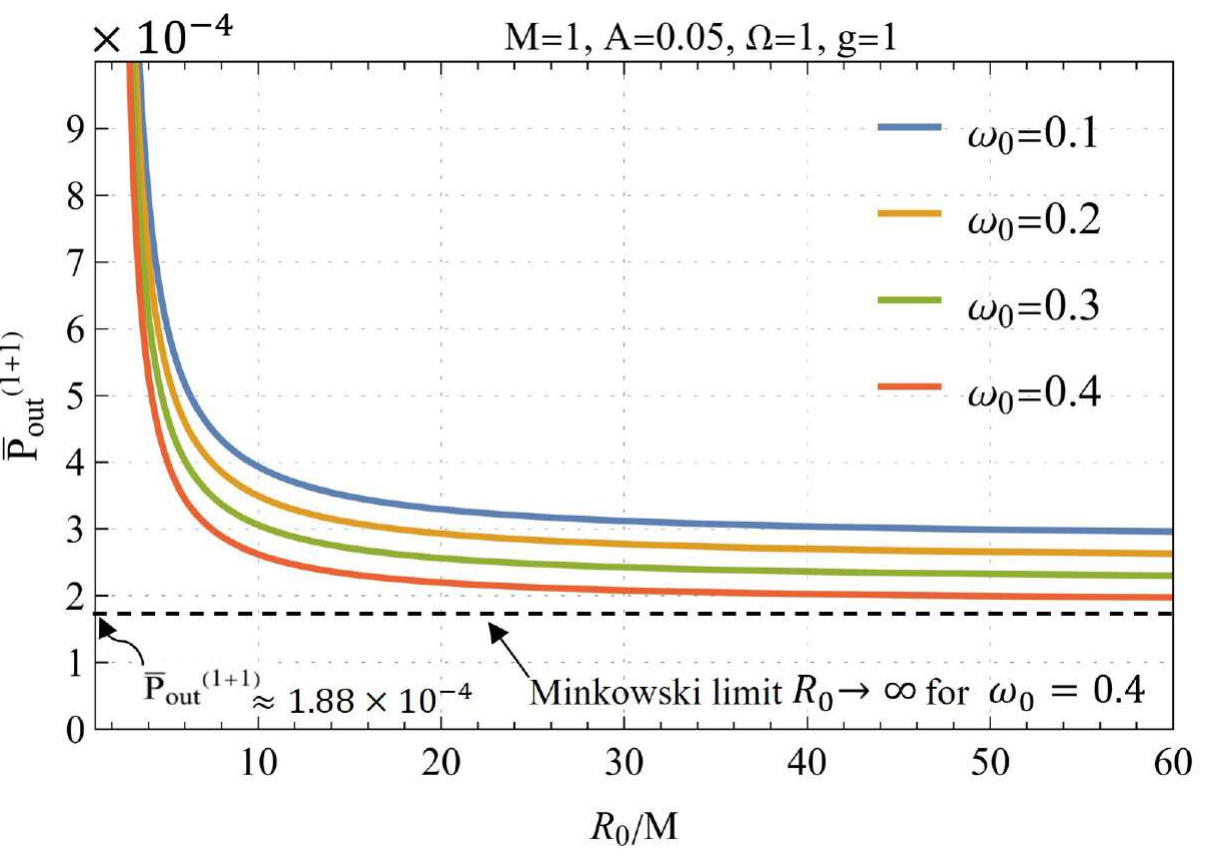}
    \caption{Total Floquet transition rate from $\overline{P}_{\text{out}}^{(1+1)}$ (Eq. \eqref{eq:totalFR}) vs normalized mean radius $R_{0}/M$ for two-level vibrating detector in Schwarzschild spacetime. The prefactor $\propto f^{-1/2}$ due to gravitational redshift and the Bessel argument $(n\Omega-\omega_{0})A\,f^{-1/2}$ lead to enhanced particle emission and smooth approach to the Minkowski limit as $R_{0}\rightarrow \infty$. We sum over integer harmonics obeying  $n\in [\omega_{0}/\Omega,25]$ for all plots consistent with the emission selection rule  $n\Omega>\omega_{0}$.}
\label{fig:PoutvsR0}%
\end{figure}
\begin{figure*}
    \centering
    \includegraphics[width=\textwidth]{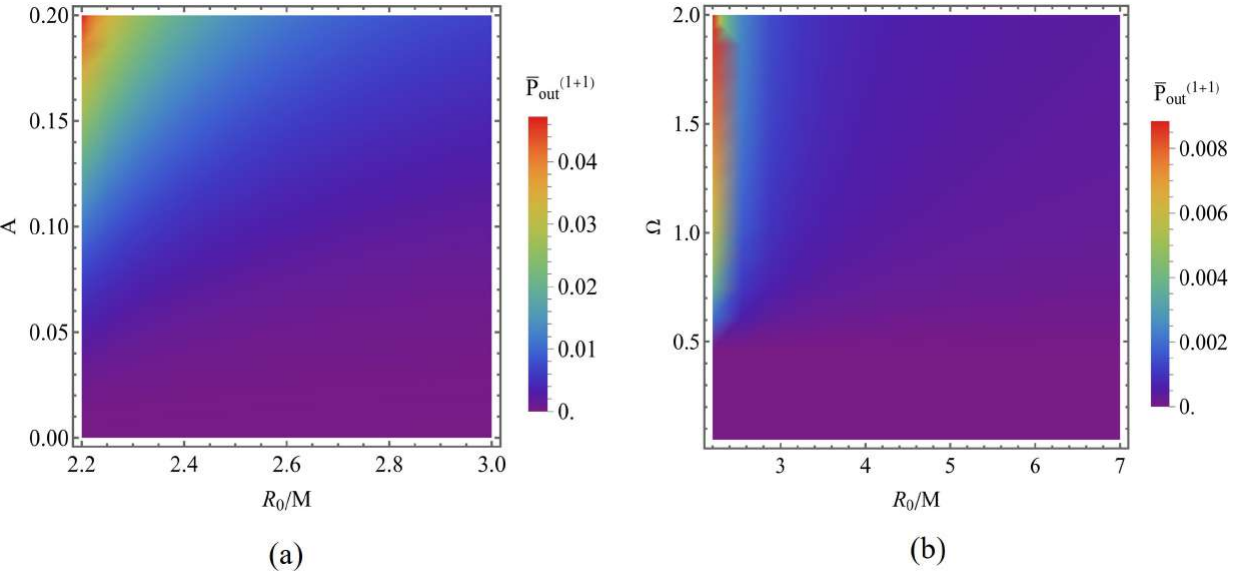}
    \caption{Total Floquet transition rate $\overline{P}^{(1+1)}_{\rm out}$ from Eq. \eqref{eq:totalFR}: (a)  $(R_{0},A)$ with $\omega_{0}=0.3, \Omega=0.4$ , and (b) $(R_{0},\Omega)$ with $\omega_{0}=0.3, A=0.05$. Colors show the cumulative contribution 
    $\Sigma_{n\geq n_{\text{min}}}\overline {P}_n^{(1+1)}$ with $n_{\text{min}}=\omega_{0}/\Omega+1$. The results tend to Minkowskian limit as $R_{0}\rightarrow \infty$ via $f(R_{0})=1-R_{0}/2M$. The redshift enhanced rate is clearly manifested in the near-horizon regime with $R_{0}\rightarrow 2M$.}
\label{fig:Pout2+1DP}%
\end{figure*}

To help appreciate the impact of various parameters on particle emission spectrum, we graphically represent total Floquet transition rate in $(1+1)$ dimensions $\overline{P}_{\text{out}}^{(1+1)}$ as follows. At the outset, Fig. \ref{fig:PoutvsR0} depicts $\overline{P}_{\text{out}}^{(1+1)}$ as a function of normalized mean distance $R_{0}$ that the atom is held fixed at from the center of the black hole while fixing the atomic transition frequency $\omega_{0}$, black hole mass $M$, oscillation amplitude $A$,  drive frequency $\Omega$, and coupling constant $g$. The plots depict (not strictly) a thermal Bose-Einstein-type profile with the total transition rate decaying as the distance from the black hole increases. The enhanced emission rate stems from prefactor $1/\sqrt{f(R_{0})}$ and a larger Bessel weight $\propto 1/\sqrt{f(R_{0})}$ with smaller $\omega_{n}$. 
This behavior clearly shows that the enhancement of particle emission intensity occurs due to the gravity of the black hole compared to the residual Minkowski limit to which the plots saturate with $f(R_{0})\rightarrow 1$ as $R_{0}/M  \rightarrow \infty$. The plots for various values of $\omega_{0}$  for the same values of $R_{0}/M$ differ in magnitude (lesser for higher $\omega_{0}$, and vice versa), which owes its origin to the energy conservation principle: detectors with higher transition frequencies are hard to be excited, and this is line with the typical Unruh dynamics. Another important point concerns the diverging tendency of the transition rate as the atom's distance from the black hole gets closer to the event horizon radius. From Eq. (\ref{eq:per_harmonic}), we can see that as $R_{0}\rightarrow 2M$, $\left(1/\omega_{n}\right) J_{n}(z)$ does not yield a well-defined limit. We may translate this as being reminiscent of the typical pathological behavior of the Boulware vacuum near the horizon. However, for the numerical consistency and for a well-defined rate near the horizon, we chose $R_{\text{min}}=2.2M$, which is in line with our working assumptions. 

We further elucidate the situation by using density plots for the total transition rate in $(1+1)$ dimensions. In Fig.\ref{fig:Pout2+1DP}(a), the total transition rate is plotted against the amplitude of the oscillations $A$ and the normalized mean distance $R_{0}/M$. One sees the much enhanced quantum effects for larger amplitude of the oscillations and/or smaller mean distances. This should not be difficult to grasp, as it simply refers to the effects of pumping more energy into the vibrating atom in the former case and the proximity of the detector to the black hole horizon in the latter case. One sees similar features in Fig.\ref{fig:Pout2+1DP}(b) where the transition rate is plotted as a function of drive frequency $\Omega$ and $R_{0}/M$. For smaller $\Omega$, the particle emission rate is negligible, no matter what the distance of the detector from the black hole. Hence, there is a threshold drive frequency that the detector needs to ensure for its excitation.

For computing the transition rate, the Bessel weight $J_{n}$ is summed over a range of harmonics $n$. Since $n$ is an integer, and for the consistency of the condition $n\Omega>\omega_{0}$,  we chose it in a way such that the minimum value of it $n_{\text{min}}=\omega_{0}/\Omega+1$, and the maximum was taken to be $n_{\text{max}}=25$. This range of $n$ suffices to produce the plots presented above as only few harmonics practically matter for significant transition rates. This is also reflected from sideband weight $J_{n}^2\left((n\Omega-\omega_{0})A/\sqrt{f(R_{0})}\right)$ which for small oscillation amplitude and higher harmonics are strongly suppressed as $J_{n}(z)\sim (z/2)^n/n!$, leaving only first few harmonic contributions.

\section{Consistency and limits} \label{sec4}
In this section, we test the Floquet transition rate's closed form derived in Section \ref{sec3.3} against several nontrivial limits and consistency checks. We also delineate the regime of validity of the approximations used in the (1+1) reduction and discuss the scope of our results in the presence of the Boulware singularity. Unless stated otherwise we keep $\hbar=c=1$, the Schwarzschild lapse $f(R)=1-2M/R$, and the detector parameters of Sections \ref{sec2}-\ref{sec3}.

Recall the per-harmonic, per-period probability Eqs. \eqref{eq:per_harmonic}-\eqref{eq:per-period}. In the asymptotically flat region $R_0\to\infty$, we have $f(R_0)\to1$. Then Eq. \eqref{eq:per-period} becomes
\begin{equation}
    \overline{P}^{(1+1)}_n\;\rightarrow{\,R_0\to\infty\,}\;
\frac{g^2}{2\,\Omega\,[\,n\Omega-\omega_0\,]}\;
J_n^{\,2}\!\left(A[\,n\Omega-\omega_0\,]\right),
\end{equation}
where $n\Omega>\omega_0$. In the limit $R_0\to\infty$ (so $f(R_0)\to1$), the result coincides with the familiar vibrating-detector resonance/Bessel-weight structure in flat spacetime; the $\Theta(n\Omega-\omega_0)$ selection rule remains implicit in the requirement $\omega>0$ for the emitted quantum (see, e.g., \cite{DeWitt:1979,Crispino:2008}).

A useful parametric rewrite highlights the energy balance in Minkowski spacetime. Defining $\Delta_n=n\Omega-\omega_0>0$ and $\xi_n=A\Delta_n$, we may write
\begin{equation}
    \overline{P}^{\text{(flat)}}_n=\frac{g^2}{2\,\Omega\,\Delta_n}\,J_n^{\,2}(\xi_n),
\end{equation}
so that higher harmonics are governed entirely by the Bessel weights $J_n(\xi_n)$ and the $1/\Delta_n$ resonance denominator. As expected, for fixed $\omega_0,\Omega$, the spectrum is concentrated around the lowest admissible $n$.

For small-argument Bessel expansion (weak driving / small amplitude), suppose the dimensionless argument of the Bessel function is small,
\begin{equation} \label{eq:small-arg_bessel}
    \xi_n(R_0)=\frac{A}{\sqrt{f(R_0)}}\,[\,n\Omega-\omega_0\,]\ll 1.
\end{equation}
Using the standard small-argument expansion,
\begin{equation} \label{eq:J_n}
    J_n(\xi)=\frac{1}{n!}\left(\frac{\xi}{2}\right)^n+\mathcal{O}(\xi^{n+2}),
\end{equation}
we obtain the leading-order scaling of each harmonic:
\begin{align}
    \overline{P}^{(1+1)}_n
&=\frac{g^2}{2\,\Omega}\,\frac{1}{\sqrt{f(R_0)}\,[\,n\Omega-\omega_0\,]}\; \nonumber \\
& \times \frac{1}{(n!)^2}\left(\frac{\xi_n(R_0)}{2}\right)^{\!2n}
\left[1+\mathcal{O}(\xi_n^2)\right].
\end{align}
Equivalently,
\begin{align}
    \overline{P}^{(1+1)}_n
&=\frac{g^2}{2\,\Omega}\,
\frac{1}{\sqrt{f(R_0)}}\,
\frac{1}{(n!)^2}\left(\frac{A}{2\sqrt{f(R_0)}}\right)^{\!2n} \nonumber \\
& \times \left[n\Omega-\omega_0\right]^{2n-1}
\left[1+\mathcal{O}(\xi_n^2)\right].
\end{align}
Let
\begin{equation}
    n_{\min} \equiv \min\{n\in\mathbb{N}: n\Omega>\omega_0\}=\big\lfloor \omega_0/\Omega \big\rfloor+1
\end{equation}
denote the first open harmonic.
Using Eq. \eqref{eq:per-period} with
$\omega_n=\sqrt{f(R_0)}(n\Omega-\omega_0)$ and
Eq. \eqref{eq:J_n}, we obtain the explicit weak-drive scaling
\begin{equation}
\overline{P}^{(1+1)}_n \;\sim\;
\frac{g^2}{2\,\Omega\,\omega_n}\,
\frac{1}{(n!)^2}\,
\left[\frac{(n\Omega-\omega_0)\,A}{2\,\sqrt{f(R_0)}}\right]^{\!2n}\, ,
\end{equation}
where $n\ge n_{\min}$. This makes transparent the competition between the $f(R_0)^{-n}$ that enters via the Bessel argument and the explicit $1/\omega_n\propto f(R_0)^{-1/2}$ prefactor.
The leading contribution at weak drive comes from $n=n_{\min}$, with higher-$n$ channels factorially suppressed.

Based from these, three points are immediate: (i) For $\xi_n\ll1$, $\overline{P}_n$ decays rapidly with $n$ due to the factorial suppression $(n!)^{-2}$ and the small parameter $A/\sqrt{f(R_0)}$, which is an indication of harmonic hierarchy; (ii) The dominant term is the first kinematically allowed harmonic (first open channel), i.e., the smallest $n$ with $n\Omega>\omega_0$. In particular, if $\omega_0<\Omega$, the $n=1$ line dominates with
\begin{align}\nonumber
    \overline{P}^{(1+1)}_{1}&\approx
\frac{g^2}{2\,\Omega}\,
\frac{1}{\sqrt{f(R_0)}}\,
\left(\frac{A}{2\sqrt{f(R_0)}}\right)^{\!2}\,[\,\Omega-\omega_0\,]^{1}\\
&=\frac{g^2 A^2}{8\,\Omega\,f(R_0)^{3/2}}\,[\,\Omega-\omega_0\,];
\end{align}
(iii) Finally, for fixed $\Omega,\omega_0,A$, decreasing $f(R_0)$ (moving inward) increases $\xi_n\propto 1/\sqrt{f(R_0)}$, tending to populate higher harmonics, but the overall prefactor carries an explicit $f(R_0)^{-1/2}$. The net behavior is controlled by the competition between these factors; the small-argument regime Eq. \eqref{eq:small-arg_bessel} quantifies when higher-$n$ channels remain negligible.

Now, we look at the static-detector and perform zero-drive checks. There are two closely related consistency checks: (i) vanishing amplitude $A\to0$ at fixed $\Omega$, and (ii) vanishing drive frequency $\Omega\to0$ at fixed $A$. For the vanishing amplitude, we use $J_n(0)=0$ for all $n\ge 1$. Then, Eq. \eqref{eq:per-period} gives
\begin{equation}
    \lim_{A\to 0}\overline{P}^{(1+1)}_n=0,\quad n\ge1.
\end{equation}
Thus, a static detector in the Boulware vacuum exhibits no period-averaged excitations in the (1+1) model away from the horizon, as expected. For the zero-frequency drive, the selection rule requires $n\Omega>\omega_0$, which cannot be satisfied for any finite $n$ if $\Omega\to0^+$ at fixed $\omega_0>0$. Hence
\begin{equation}
    \lim_{\Omega\to0^+}\overline{P}^{(1+1)}_n=0 \quad \text{for all fixed } n,
\end{equation}
and the total Floquet transition rate vanishes. This is consistent with the fact that slow, adiabatic deformations of the worldline cannot resonantly bridge a fixed gap $\omega_0$. In terms of consistency with energy balance, in either limit, the emitted Killing frequency in Eq. \eqref{eq:killing_freq} becomes non-positive or the Bessel weight vanishes, preventing unphysical negative-frequency radiation.

Next, we explore the regime of validity near the horizon. Here we make the constraints quantitative when $R_0$ approaches the horizon $2M$ so that $f(R_0)\ll1$. For the nonrelativistic radial motion, the proper-time velocity bound
\begin{equation} \label{eq:prop-time_vel_bound}
    |\dot r(\tau)|\sim A\Omega\ll \sqrt{f(R_0)},
\end{equation}
ensures that redshift and tortoise-coordinate factors may be evaluated at $R_0$ and that higher-order Doppler corrections are negligible. Solving Eq. \eqref{eq:prop-time_vel_bound} for $A$ gives
\begin{equation}
    A \ll \frac{\sqrt{f(R_0)}}{\Omega}.
\end{equation}
Thus as $R_0\to 2M$ (so $f(R_0)\to0$), the admissible amplitude shrinks linearly with $\sqrt{f(R_0)}$. For the frequency window and separation of scales, the resonance width of the single-period sinc kernel is $\Delta\omega\sim \Omega/\sqrt{f(R_0)}$, while the resonant frequency scales as $\omega_n\sim \sqrt{f(R_0)}\,(n\Omega-\omega_0)$. To avoid overlap of adjacent harmonics, one requires
\begin{equation}
    (n+1)\Omega-\omega_0-(n\Omega-\omega_0)=\Omega \gg \text{intrinsic broadening},
\end{equation}
which is automatically satisfied at the level of our period-averaged treatment (Section \ref{sec3.2}). In perturbation theory, additional broadening $\propto g^2$ is assumed to be small. If one wishes to work in the small-argument Bessel expansion regime, in addition to asymptotically flat region approximation, then Eq. \eqref{eq:small-arg_bessel} implies the stronger bound
\begin{equation}
    \frac{A}{\sqrt{f(R_0)}}\,[\,n\Omega-\omega_0\,]\ll1.
\end{equation}
For the first open channel $n=n_{\min}$ with $n_{\min}\Omega>\omega_0$, this becomes
\begin{equation}
    A \ll \frac{\sqrt{f(R_0)}}{n_{\min}\Omega-\omega_0}.
\end{equation}
Again, the allowed amplitude collapses as $\sqrt{f(R_0)}$ near the horizon. For the validity of the (1+1) reduction, the s-wave (1+1) model neglects the (3+1) Regge–Wheeler potential. Close to the horizon, greybody effects are subleading for purely outgoing modes in the (1+1) kinematics, but for quantitative predictions near $R_0\sim 2M$ one should ultimately include the barrier (Section \ref{sec5}). Our results remain reliable as long as the relevant wavelengths $\lambda_n\sim 1/\omega_n$ are large compared with the curvature scale and the effective potential varies slowly across a wavelength in the region probed by the worldline.

Collecting all these results near the horizon, a conservative near-horizon validity domain is $(f(R_0)\gg \varepsilon)$
\begin{equation} \label{eq:validity-domain}
    A\Omega \ll \sqrt{f(R_0)},\quad
    A \ll \min\!\left\{\frac{\sqrt{f(R_0)}}{\Omega},\ \frac{\sqrt{f(R_0)}}{n_{\min}\Omega-\omega_0}\right\},
\end{equation}
where $\varepsilon$ is a problem-dependent threshold beyond which one must account for both Boulware singular behavior and (3+1) scattering.

We make some final remarks about the Boulware singularity. The Boulware vacuum $|0\rangle_B$ is defined to be empty with respect to the static Killing time at infinity and is regular there, but the renormalized stress tensor diverges as $R\to 2M$. Our observable is a transition probability computed to leading order in $g$ along a timelike worldline at fixed mean radius $R_0>2M$, with period averaging performed over a finite (single-cycle) window. Within this scope, any fixed $R_0>2M$ with $f(R_0)>0$, Eqs. \eqref{eq:per_harmonic}–\eqref{eq:per-period} are finite. The only explicit dependence on $f(R_0)$ is through the factors $\omega_n=\sqrt{f(R_0)}(n\Omega-\omega_0)$ and the Bessel argument $\xi_n=A[n\Omega-\omega_0]/\sqrt{f(R_0)}$. As $R_0\to 2M$, the dimensionless argument $\xi_n$ grows like $f(R_0)^{-1/2}$, while the prefactor scales like $f(R_0)^{-1/2}$ (see Eq. \eqref{eq:per-period}). However, our kinematic bound Eq. \eqref{eq:prop-time_vel_bound} forces $A$ to scale at most as $\sqrt{f(R_0)}/\Omega$, which in turn prevents $\xi_n$ from diverging:
\begin{equation}
    \xi_n \lesssim \frac{A}{\sqrt{f(R_0)}}(n\Omega-\omega_0)
\;\lesssim\; \frac{n\Omega-\omega_0}{\Omega}=\mathcal{O}(1).
\end{equation}
Hence, within the perturbative, nonrelativistic regime, $\overline{P}^{(1+1)}_n$ does not blow up solely due to the $f(R_0)\to0$ limit; instead, the range of admissible amplitudes collapses and the (1+1) model ceases to be predictive before encountering the formal Boulware divergence in local stress tensor observables.

We do not extrapolate Eqs. \eqref{eq:per-period}–\eqref{eq:per-period} into the regime where Eq. \eqref{eq:prop-time_vel_bound} fails, nor do we assert regularity for local geometric observables (e.g., $\langle T_{ab}\rangle$) along the worldline. Our calculation is a first-order response functional, insensitive to the near-horizon energy density pathologies of $|0\rangle_B$ as long as the worldline remains at fixed $R_0>2M$ and the motion is sufficiently mild. A full treatment arbitrarily close to the horizon requires switching to the Unruh/Hartle–Hawking states and (3+1) propagation with back-reaction beyond our scope.

\section{(3+1) outlook} \label{sec5}
The (1+1) result of Section \ref{sec3} encapsulates the essential kinematics (redshift, resonance, period averaging). In (3+1) dimensions, two new ingredients enter: (i) angular momentum decomposition with spherical harmonics on the 2–2-sphere, and (ii) propagation through the Schwarzschild scattering potential (greybody factors). To leading order in the coupling $g$, the structure of Floquet resonances \cite{magnus2013hill} and the universal period factor $2\pi/\Omega$ persist unchanged; the principal modifications appear in the mode basis, normalization, and a transmission factor $|\mathcal{T}_{\omega\ell}|^2$ that weights each partial wave.

A convenient (3+1) mode basis for a massless scalar on the Schwarzschild geometry is
\begin{align}
    u_{\omega\ell m}^{\rm out}(t,r,\theta,\phi)
=\frac{1}{\sqrt{4\pi\omega}}\;\frac{\psi_{\omega\ell}(r)}{r}\;Y_{\ell m}(\theta,\phi)\;e^{-i\omega t},
\end{align}
with $\omega>0,\ \ell\in\mathbb{N}_0,\ m=-\ell,\dots,\ell$
and  radial wavefunction $\psi_{\omega\ell}$ solving the Regge–Wheeler equation for $f(r)=1-\frac{2M}{r}$ is
\begin{equation} \label{eq:reg_wheel}
    \frac{d^2\psi_{\omega\ell}}{dr_*^2}
+\left[\omega^2 - V_\ell(r)\right]\psi_{\omega\ell}=0,
\end{equation}
where the scattering potential $V_\ell(r)=f(r)\!\left(\frac{\ell(\ell+1)}{r^2}+\frac{2M}{r^3}\right)$.
It is worth noting that for the s-wave ($\ell=0$), the potential becomes $V_0(r) = f(r)(2M/r^3)$, a term that was neglected in the (1+1) model for simplicity but is fully accounted for in the (3+1) greybody factor $|\mathcal{T}_{\omega_0}|^2$ \cite{Chandrasekhar:1985kt,Frolov_1998}. Asymptotically, $\psi_{\omega\ell}\sim A_{\omega\ell}^{\rm in}e^{-i\omega r_*}+A_{\omega\ell}^{\rm out}e^{+i\omega r_*}$ at $r_*\to+\infty$; the greybody transmission is $|\mathcal{T}_{\omega\ell}|^2=|A_{\omega\ell}^{\rm out}|^2/|A_{\omega\ell}^{\rm hor}|^2$, with $A_{\omega\ell}^{\rm hor}$ the unit-amplitude ingoing wave at the horizon. The Unruh–DeWitt interaction along a worldline $x(\tau)=(t(\tau),r(\tau),\theta(\tau),\phi(\tau))$ then yields (schematically) the first-order amplitude
\begin{align}
    \mathcal{A}_{\omega\ell m} &\propto \frac{g}{\sqrt{4\pi\omega}}
\int d\tau\,e^{+i\omega_0\tau}\;
\frac{\psi_{\omega\ell}(r(\tau))}{r(\tau)} \nonumber \\
&\times \,Y_{\ell m}(\theta(\tau),\phi(\tau))\,
e^{-i\omega\,t(\tau)}.
\end{align}

For the radial small-amplitude trajectory used in Section \ref{sec3},
\begin{equation}
    r(\tau)=R_0+A\cos(\Omega\tau),\quad \theta(\tau)=\theta_0,\quad \phi(\tau)=\phi_0,
\end{equation}
\begin{figure*}
    \centering
    \includegraphics[width=\textwidth]{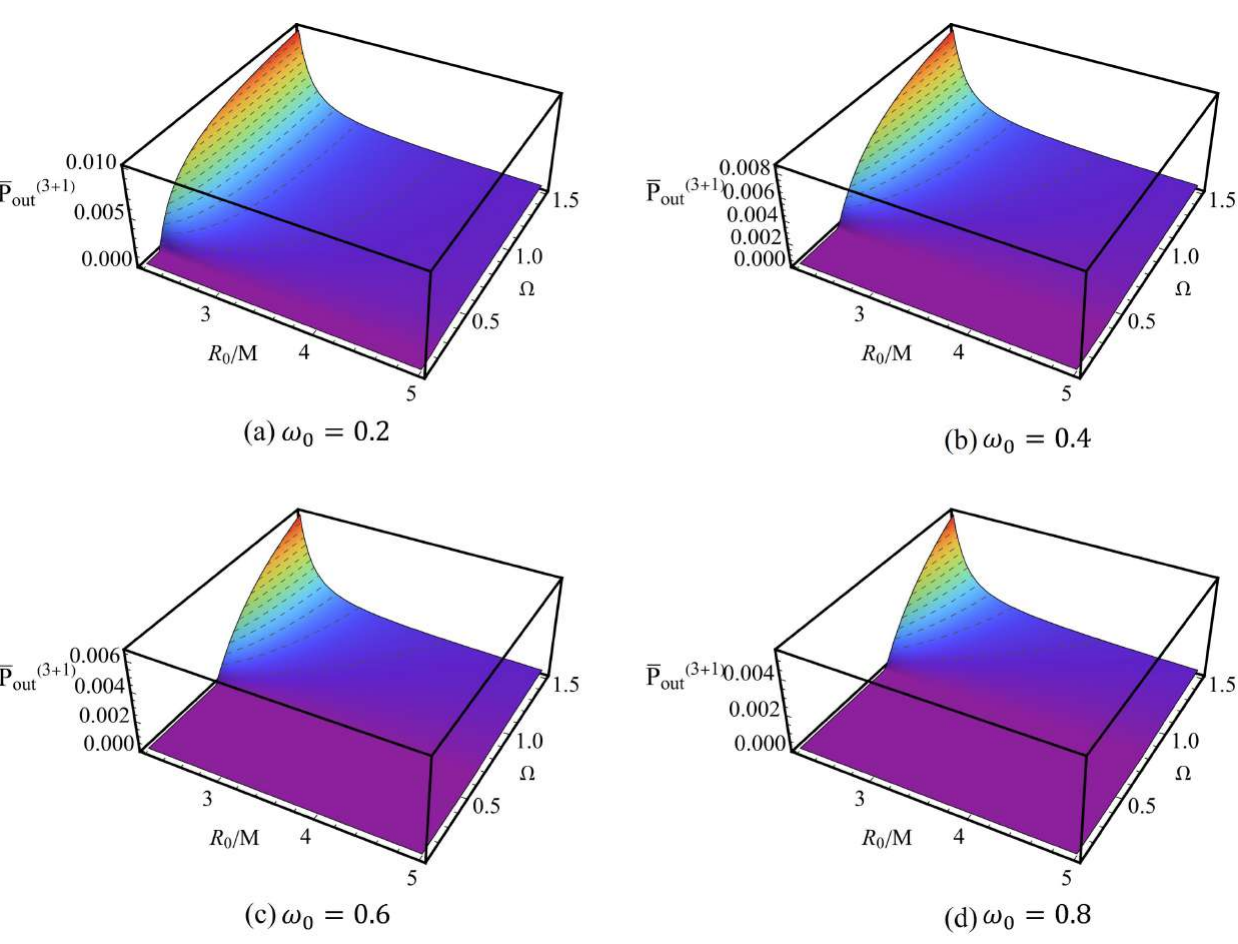}
    \caption{$(3+1)$-dimensional total Floquet transition rate $\overline{P}_{\text{out}}^{(3+1)}$ as a function of normalized mean distance $R_{0}/M$ and drive frequency $\Omega$ for two-level vibrating atom in Schwarzschild spacetime (in units $G=\hbar=c=1$). Plots represent radiative dynamics for four values of (a) $\omega_{0}=0.2$, (b) $\omega_{0}=0.4$, (c) $\omega_{0}=0.6$, (d) and $\omega_{0}=0.8$ . The chosen parameters include $M=1, g=1, \text{and}\  A=0.05$. Once again the  range for harmonics $n\in \left[\omega_{0}/\Omega,25\right]$. For all four cases, the channel opening threshold condition $\omega_{0}/n$ pushes $\Omega$ to higher values as $\omega_{0}$ increases, leading to enhanced inert zone for transition rates. Each harmonic $n$ entails its own specific $\Omega$ threshold. }
    \label{Pout-3D}%
\end{figure*}
with the nonrelativistic bound $A\Omega\ll\sqrt{f(R_0)}$, the angular dependence reduces to a fixed spherical harmonic $Y_{\ell m}(\theta_0,\phi_0)$. Moreover, the same Jacobi–Anger expansion that generated Floquet harmonics in (1+1) applies to the phase $\omega t(\tau)-\omega_0\tau$ and to the mild radial dependence of $\psi_{\omega\ell}(r(\tau))$ about $R_0$. Retaining the leading (harmonic) content and period-averaging over a single cycle $T=2\pi/\Omega$ gives the schematic (3+1) generalization of Section \ref{sec3.3}:
\begin{align} \label{eq:3+1_gen}
	&\overline{P}^{(3+1)}_{\rm out}
	\;\simeq\;
	\sum_{n=1}^{\infty}\Theta(n\Omega-\omega_0)\; \nonumber \\
	& \times\sum_{\ell=0}^{\infty}\sum_{m=-\ell}^{\ell}
	\;\frac{2\pi}{\Omega}\,\frac{g^2}{4\pi\,\omega_n}\;
	\Big|\mathcal{S}_{\ell m}(R_0,\theta_0,\phi_0)\Big|^2
	\;|\mathcal{T}_{\omega_n\ell}|^2\;
	\mathcal{J}_n^2
\end{align}
with
\begin{align}\nonumber
    \omega_n&=\sqrt{f(R_0)}\,[\,n\Omega-\omega_0\,]\\
\nonumber\mathcal{S}_{\ell m}& =\frac{\psi_{\omega_n\ell}(R_0)}{R_0}\,Y_{\ell m}(\theta_0,\phi_0),\\
\nonumber \mathcal{J}_n&=J_n\!\left(\frac{(n\Omega-\omega_0)\,A}{\sqrt{f(R_0)}}\right).
\end{align}
In this schematic expression, the factor \(|\mathcal{T}_{\omega_n\ell}|^2\) is the usual Schwarzschild greybody factor for each partial wave, encoding the transmission probability through the Regge-Wheeler potential \(V_\ell(r)\) \cite{Birrell:1982ix,Wald:1995yp}. The (1+1)-dimensional result of Section \ref{sec3} is recovered by restricting to the s-wave sector \(\ell=0\), for which the potential barrier is weakest and \(|\mathcal{T}_{\omega 0}|^2\to 1\) at high frequency, and by suppressing the angular dependence in \(\mathcal{S}_{\ell m}\). Higher-\(\ell\) modes are filtered by the barrier and thus give subleading corrections to the total outgoing flux for a detector located not too far from the peak of \(V_\ell(r)\).

We plot the $3+1$-dimensional total transition rate  $\overline{P}_{\text{out}}^{(3+1)}$ given in Eq.(\ref{eq:3+1_gen}) in Fig.\ref{Pout-3D} for the indicated values of $\omega_{0}$. The figures represent results for s-wave baseline with $l=0,m=0$, and greybody factors $\left|\mathcal{T}_{\omega_n\ell}\right|^2$ set to unity. The overall nature of particle emission is not fundamentally very different from the $(1+1)$-dimensional case. The channel opening thresholds continue to hold for $\Omega=\omega_{0}/n$. Once again, the enhanced emission spectrum occurs for the near-horizon regime.

It is noteworthy that the channel thresholds at $\Omega=\omega_{0}/n$ $(n=1,2,3...)$ are independent of $R_{0}$. All four cases showcase an inert zone for transition rate with zero emission, no matter what the value of $R_{0}/M$, as long as the $\Omega$ remains below a certain threshold. The threshold values of $\Omega$ are mathematically controlled by Heaviside function. These threshold values of $\Omega$ are shifted to higher values as $\omega_{0}$ increases across all four cases, as depicted in Fig.\ref{Pout-3D} (a)-(d), expanding the low$-\Omega$ inert zone further. One should be mindful of the fact that the threshold $\Omega$ is different for each harmonic $n$. Since higher $n$ harmonics are suppressed, it is hard to appreciate those tiny values of transition rate, and plots show zero rate for most of the harmonics.  Furthermore, we note that even though gravity amplifies the transition rate for lower $R_{0}/M$, it does not shift the kinematic thresholds in $\Omega$.

Eq. \eqref{eq:3+1_gen} reduces exactly to the (1+1) expression Eq. \eqref{eq:per_harmonic} when (i) the s-wave dominates, $\ell=0$ (so $Y_{00}=1/\sqrt{4\pi}$, $\psi_{\omega 0}(r)\propto e^{+i\omega r_*}$ outside the barrier), and (ii) the barrier is neglected so that $|\mathcal{T}_{\omega 0}|^2\to 1$, $|\psi_{\omega 0}(R_0)|\to 1$. In general, $|\mathcal{T}_{\omega\ell}|^2$ suppresses higher $\ell$ at low frequencies and encodes curvature-induced backscattering. For purely radial motion at fixed angles, the dominant magnetic number is $m=0$ (by parity and selection at small velocity), so that, for the leading contribution,
\begin{equation}
    \sum_{m=-\ell}^{\ell}\big|Y_{\ell m}(\theta_0,\phi_0)\big|^2 \;\to\; \big|Y_{\ell 0}(\theta_0)\big|^2
\end{equation}
while for isotropic angle averages one may replace $|Y_{\ell m}|^2$ by $(2\ell+1)/(4\pi)$.

Two practical routes suggest themselves for quantitative (3+1) estimates: (i) For $\omega R_0\ll1$, use the known asymptotics of $V_\ell(r)$ to approximate $|\mathcal{T}_{\omega\ell}|^2$ (exponentially small for $\ell\ge1$), confirming s-wave dominance and giving controlled corrections to Eq. \eqref{eq:per_harmonic}; (ii) Compute $|\mathcal{T}_{\omega\ell}|^2$ by integrating Eq. \eqref{eq:reg_wheel} outward from the horizon with purely ingoing boundary conditions, extract $A_{\omega\ell}^{\rm out}$ at large $r_*$, and insert into Eq. \eqref{eq:3+1_gen}. The Floquet/Bessel factor $\mathcal{J}_n$ and the universal $2\pi/\Omega$ remain as in (1+1).

As a final remark, the state choice in (3+1) may still be Boulware, Unruh, or Hartle–Hawking vacua. The derivation above mirrors our (1+1) Boulware analysis; switching to Unruh/Hartle–Hawking amounts to replacing the mode occupations and adding stimulated terms in the standard way, without altering the resonant skeleton of Eq. \eqref{eq:3+1_gen}. This provides a clear path from our baseline (1+1) formulas to phenomenology at $\mathscr{I}^+$ with curvature-induced filtering by $|\mathcal{T}_{\omega\ell}|^2$.

\section{Conclusion} \label{sec6}
We have developed a first-principles, mode-amplitude derivation of the Floquet transition rate for a linearly coupled Unruh–DeWitt detector executing small-amplitude radial oscillations in the exterior Schwarzschild geometry. Working within an s-wave (1+1) reduction and the Boulware vacuum, we showed that the finite-time window integral over a single oscillation cycle cleanly resolves the Floquet spectrum of the driven worldline and produces a universal period factor $2\pi/\Omega$. The particle emission spectrum is analogous to a thermal Bose-Einstein-like spectrum with a suppressed tail for asymptotic distances of the atom from the black hole. The ill-defined spectrum at the black hole horizon highlights the pathological character of the Boulware vacuum. The resulting closed form (Section \ref{sec3.3}), given by Eq. \eqref{eq:per_harmonic}, encapsulates the interplay between gravitational redshift, kinematic selection ($n\Omega>\omega_0$), and harmonic content controlled by Bessel weights $J_n$. The derivation highlights two conceptual points: (i) the emergence of resonant delta-comb structure from the finite-time sinc kernel, and (ii) the fact that period averaging is the natural object for comparing across switching profiles, as it divides out the trivial growth with observation time while preserving resonant areas.

The consistency analysis in Section \ref{sec4} confirms that our expression reproduces the expected Minkowski limit, exhibits the factorial hierarchy of small-argument Bessel asymptotics, and vanishes continuously in the static and adiabatic limits. Near the horizon, we identified a kinematic regime ($A\Omega\ll \sqrt{f(R_0)}$) within which the method remains predictive: as $R_0\to 2M$, the allowed amplitude window collapses before the Boulware stress-tensor pathology becomes operational for our observable, thereby preventing unphysical divergences in $\overline{P}^{(1+1)}_n$ within the validity domain. This clarifies the scope of the (1+1) Boulware analysis and delineates where a change of state (Unruh/Hartle–Hawking) or a more complete (3+1) treatment must take over.

While our main result is formulated in (1+1) dimensions, Section \ref{sec5} sketches the (3+1) generalization in which the resonant skeleton (the Floquet selection rule, period factor $2\pi/\Omega$, and Bessel weights) survives intact, but each partial wave is filtered by the Schwarzschild barrier through a greybody transmission $|\mathcal{T}_{\omega\ell}|^2$. In this sense, curvature influences the spectrum predominantly through (a) redshift at the worldline (entering $\omega_n$ and the Bessel argument) and (b) propagation from the source region to $\mathscr{I}^+$ (entering $|\mathcal{T}_{\omega\ell}|^2$). This separation suggests a practical program: compute the local, redshifted source spectrum using our closed form, then convolve with transmission probabilities to obtain observable fluxes.

Methodologically, an important contribution of this work is to place the mode-amplitude route, which is often viewed as complementary to the Wightman-function response functional, on equal footing for time-periodic trajectories. The Jacobi–Anger–to–delta-comb chain makes explicit how resonance, not stationarity, underlies detector excitation in curved backgrounds. Appendix \ref{appendixA} will provide the distributional foundation for the sinc $\to$ Dirac-delta limit under various switchings; an optional Appendix \ref{appendixB} can show the equivalence of our mode-amplitude expression with the response-functional route, strengthening the robustness of the result.

There are several natural extensions. First, completing the (3+1) program with numerical Regge–Wheeler integration would quantify greybody suppression across $\ell$ and test s-wave dominance as a function of $R_0$, $\Omega$, and $A$. Second, changing the quantum state of the field (Unruh or Hartle–Hawking) introduces stimulated terms without altering the resonant backbone, enabling applications to near-horizon steady fluxes and black-hole thermality. Third, going beyond radial oscillations to circular or epicyclic motion (and, in Kerr, to frame-dragging-locked orbits) would probe selection rules tied to azimuthal quantum numbers and superradiant windows. Fourth, incorporating finite spatial smearing of the detector and higher-order perturbative effects would address line broadening, level shifts, and backreaction. Finally, optimizing trajectories for maximal spectral weight at selected harmonics suggests a control-theoretic angle on curvature-assisted radiation engineering.

From a broader perspective, our closed-form expression for the redshifted Floquet spectrum of an oscillating detector provides a simple analytic laboratory for testing how curvature modifies Unruh-like effects. Because the result reduces smoothly to the known flat-space vibrating-atom limit while encoding the gravitational potential only through \(f(R_0)\) and greybody factors, it suggests a clear strategy for exporting tabletop acceleration-radiation proposals to astrophysically relevant spacetimes. We expect that the formal tools developed here, especially the finite-time Floquet analysis in curved space, will be useful in future studies of more realistic detectors, different quantum states (Unruh/Hartle-Hawking), and dynamical geometries.

\acknowledgments
R. P. and A. O would like to acknowledge networking support of the COST Action CA21106 - COSMIC WISPers in the Dark Universe: Theory, astrophysics and experiments (CosmicWISPers), the COST Action CA22113 - Fundamental challenges in theoretical physics (THEORY-CHALLENGES), the COST Action CA21136 - Addressing observational tensions in cosmology with systematics and fundamental physics (CosmoVerse), the COST Action CA23130 - Bridging high and low energies in search of quantum gravity (BridgeQG), and the COST Action CA23115 - Relativistic Quantum Information (RQI) funded by COST (European Cooperation in Science and Technology). R. P. and A. O. would also like to acknowledge the funding support of SCOAP3.

\appendix
\section{Distributional and finite-time details} \label{appendixA}
This appendix gathers the Fourier–analytic statements that underlie the steps used in Sections \ref{sec3.1}-\ref{sec3.2}: (i) how finite-time windows produce sinc kernels; (ii) how either one very long window or many repeated periods generate Dirac deltas (or Dirac combs) in the sense of tempered distributions; (iii) how rectangular and smooth switchings compare; and (iv) how the resonance constraint $\alpha(\omega)+m\Omega=0$ yields $\delta(\omega-\omega_n)$ with the correct Jacobian.

We work in the space of tempered distributions $\mathcal{S}'(\mathbb{R})$, using the unitary convention \cite{Wald:1995hf,Wald:1995yp}
\begin{align} \label{eq:A1}
    &\widehat{f}(\beta)=\int_{-\infty}^{\infty}\! d\tau\; e^{+i\beta\tau}\,f(\tau),\nonumber \\
&f(\tau)=\frac{1}{2\pi}\int_{-\infty}^{\infty}\! d\beta\; e^{-i\beta\tau}\,\widehat{f}(\beta).
\end{align}

\subsection{A.1 Single finite window \texorpdfstring{$\Rightarrow$}{} sinc kernel} \label{secA.1}
Let the rectangular (boxcar) window of duration $T>0$ centered at $\tau=0$ be
\begin{equation} \label{eq:A2}
    w_T(\tau)=\mathbf{1}_{[-T/2,T/2]}(\tau)=
\begin{cases}
	1, & |\tau|\le T/2,\\
	0, & \text{otherwise}.
\end{cases}
\end{equation}
Its Fourier transform is
\begin{equation} \label{eq:A3}
    \widehat{w}_T(\beta)=\int_{-T/2}^{T/2}\! d\tau\, e^{+i\beta\tau}
= e^{+i\beta T/2}\,T\,\mathrm{sinc}\!\left(\frac{\beta T}{2}\right),
\end{equation}
where $\mathrm{sinc}(x)= \sin x / x$, which is the origin of the sinc kernel used in Section \ref{sec3.1} (Eq. \eqref{eq:sing_per_A} there). If the time-domain integrand contains a phase $e^{+i\beta\tau}$ with $\beta$ possibly depending on $\omega$ and integer labels, then multiplying by $w_T$ in $\tau$ multiplies the frequency-domain amplitude by $\widehat{w}_T(\beta)$, i.e. by $e^{+i\beta T/2}T\,\mathrm{sinc}(\beta T/2)$.

\subsection{A.2 Distributional limit: \texorpdfstring{$T\,\mathrm{sinc}\!\left(\frac{\beta T}{2}\right)\to 2\pi\,\delta(\beta)$}{}} \label{secA.2}
Proposition A.1 (sinc $\to \delta$). In $\mathcal{S}'(\mathbb{R})$,
\begin{equation} \label{eq:A4}
    \lim_{T\to\infty} e^{+i\beta T/2}\,T\,\mathrm{sinc}\!\left(\frac{\beta T}{2}\right)=2\pi\,\delta(\beta).
\end{equation}
\textit{Proof.} Let $\varphi\in\mathcal{S}(\mathbb{R})$. Consider
\begin{align}\nonumber
    \int_{-\infty}^{\infty}\! &d\beta\; \varphi(\beta)\,e^{+i\beta T/2}\,T\,\mathrm{sinc}\!\left(\frac{\beta T}{2}\right)\\
&=\int_{-T/2}^{T/2}\! d\tau\;\left[\int_{-\infty}^{\infty}\! d\beta\; \varphi(\beta)\,e^{+i\beta(\tau+T/2)}\right]\\
&=2\pi\int_{-T/2}^{T/2}\! d\tau\;\check{\varphi}(\tau+T/2) \nonumber,
\end{align}
where $\check{\varphi}$ is the inverse transform of $\varphi$ with the convention Eq. \eqref{eq:A1}. Changing variables $u=\tau+T/2$ yields
\begin{align}\nonumber
    2\pi\int_{0}^{T}\! &du\,\check{\varphi}(u)\rightarrow[T\to\infty]{}2\pi\int_{0}^{\infty}\! du\,\check{\varphi}(u)\\
    &=2\pi\,\check{\varphi}(0)\\
&=2\pi\,\frac{1}{2\pi}\int_{-\infty}^{\infty}\! d\beta\;\varphi(\beta)=2\pi\,\varphi(0), \nonumber
\end{align}  
using the Riemann–Lebesgue lemma and the fact $\check{\varphi}\in\mathcal{S}$ is integrable. Thus the limit functional is $\varphi\mapsto 2\pi\,\varphi(0)$, i.e. $2\pi\delta$. $_\square$

Some final remarks: (i) The phase factor $e^{+i\beta T/2}$ is immaterial for the limit since it equals $1$ at $\beta=0$ and only shifts the kernel without changing the distributional limit; and (ii) The statement generalizes to any window $w_T$ obtained by dilation of a fixed $w\in L^1$ with $\int w=1$; see Section \ref{secA.4}

\subsection{A.3 Many periods \texorpdfstring{$\Rightarrow$}{} Dirichlet/Fejér kernels and Dirac combs} \label{secA.3}
Consider $N\in\mathbb{N}$ consecutive periods of length $T=2\pi/\Omega$, i.e.
\begin{equation} \label{eq:A5}
    W_{N,T}(\tau)=\sum_{k=-N}^{N} w_T(\tau-kT).
\end{equation}
Its Fourier transform factorizes:
\begin{align} \label{eq:A6}
    \widehat{W}_{N,T}(\beta)&=\widehat{w}_T(\beta)\,\underbrace{\sum_{k=-N}^{N} e^{+i\beta kT}}_{D_N(\beta T)} \nonumber \\
&= e^{+i\beta T/2}T\,\mathrm{sinc}\!\left(\frac{\beta T}{2}\right)\,D_N(\beta T),
\end{align}
where
\begin{equation} \label{eq:A7}
    D_N(\theta)=\sum_{k=-N}^{N}e^{ik\theta}=\frac{\sin\!\left((2N+1)\theta/2\right)}{\sin(\theta/2)}
\end{equation}
is the Dirichlet kernel. The power kernel relevant for probabilities is the Fejér kernel,
\begin{align} \label{eq:A8}
    F_N(\theta)&=\frac{1}{2N+1}\,\big|D_N(\theta)\big|^2 \nonumber \\
&=\frac{1}{2N+1}\left(\frac{\sin\!\left((2N+1)\theta/2\right)}{\sin(\theta/2)}\right)^{\!2}.
\end{align}
As $N\to\infty$,
\begin{equation} \label{eq:A9}
    \frac{1}{2N+1}\,\big|\widehat{W}_{N,T}(\beta)\big|^2
\;\to\; (2\pi)\sum_{m\in\mathbb{Z}}\delta(\beta-m\Omega),
\end{equation}
in $\mathcal{S}'$, i.e. a Dirac comb with spacing $\Omega$. More precisely, using Poisson resummation,
\begin{equation} \label{eq:A10}
    \sum_{k\in\mathbb{Z}} e^{+i\beta kT}=\frac{2\pi}{T}\,\sum_{m\in\mathbb{Z}}\delta(\beta-m\Omega),
\end{equation}
where $\Omega=2\pi/T$ so that $\widehat{W}_{\infty,T}(\beta)\propto \widehat{w}_T(\beta)\sum_m \delta(\beta-m\Omega)$. In the probability (quadratic) level, the Fejér kernel $F_N$ converges to the comb in the sense of Cesàro means; the normalization is such that each spike integrates to $2\pi/T=\Omega$.

We interpret these results as follows: For one long observation: use Eq. \eqref{eq:A4}, for many periods at fixed $T$: use Eq. \eqref{eq:A9} or Eq. \eqref{eq:A10}. In both cases, resonances appear at the harmonic set $\beta=m\Omega$, and the area under each peak per unit observation time is independent of the detailed window profile (see also Section \ref{secA.4})

\subsection{A.4 Smooth switching: scaling limits and endcaps} \label{secA.4}
Let $\chi\in\mathcal{S}(\mathbb{R})$ be a smooth, even, compactly supported (or rapidly decaying) window with $\int d\tau\,\chi(\tau)=1$. Define the scaled window of width $T$ by
\begin{equation} \label{eq:A11}
    \chi_T(\tau)=\frac{1}{T}\,\chi\!\left(\frac{\tau}{T}\right).
\end{equation}
Then $\int \chi_T=1$ and
\begin{align} \label{eq:A12}
    \widehat{\chi_T}(\beta)&=\int\! d\tau\, e^{+i\beta\tau}\,\frac{1}{T}\chi\!\left(\frac{\tau}{T}\right) \nonumber \\
&=T\int\! du\, e^{+i\beta T u}\,\chi(u)=T\,\widehat{\chi}(\beta T).
\end{align}
As $T\to\infty$, $\widehat{\chi_T}(\beta)\to 2\pi\,\delta(\beta)$ in $\mathcal{S}'$ by the same argument as in Proposition A.1, since $\widehat{\chi}\in\mathcal{S}$ and $\widehat{\chi}(0)=2\pi$. Thus, any sufficiently regular window with unit area produces the same $2\pi\delta$ limit. Relative to the rectangular case Eq. \eqref{eq:A3}, smooth windows control ringing and side-lobes but leave the integrated weight (the delta area) invariant.

For a single period “flat-top” smooth window, let
\begin{equation} \label{eq:A13}
    \chi_{T,\epsilon}(\tau)=
\begin{cases}
	1, & |\tau|\le \frac{T}{2}-\epsilon,\\
	\text{smooth roll-on/off}, & \frac{T}{2}-\epsilon < |\tau| \le \frac{T}{2},\\
	0, & |\tau|>\frac{T}{2},
\end{cases}
\end{equation}
with $\epsilon\ll T$. Then $\widehat{\chi}_{T,\epsilon}(\beta)=e^{+i\beta T/2}T\,\mathrm{sinc}(\beta T/2)+\mathcal{O}(\epsilon)$ uniformly on compact $\beta$-sets. Hence, all period-averaged expressions differ from the rectangular case by $\mathcal{O}(\epsilon/T)$ endcap corrections, which vanish when we (i) take the large-time limit or (ii) divide by $T$ to define a per-period average. The takeaway is that the shape of the window affects only the lineshape (side-lobes, broadening); the area per line per period, the quantity that survives time averaging, is universal.

\subsection{A.5 From resonance constraints to \texorpdfstring{$\delta(\omega-\omega_n)$}{} (Jacobian)} \label{secA.5}
The resonance variable used in the main text is
\begin{equation} \label{eq:A14}
    \beta(\omega,m)=\alpha(\omega)+m\Omega, \quad 
\alpha(\omega)=\omega_0+\frac{\omega}{\sqrt{f(R_0)}}.
\end{equation}
The distributional limits Eq. \eqref{eq:A4}, Eq. \eqref{eq:A6} imply that, after period averaging, the $\beta$-dependence becomes a comb $\sum_{m\in\mathbb{Z}}\delta(\beta-m\Omega)$, or, equivalently, for each $m$,
\begin{align} \label{eq:A15}
    \delta\!\left(\alpha(\omega)+m\Omega\right)
&=\delta\!\left(\frac{\omega}{\sqrt{f(R_0)}}+\omega_0+m\Omega\right) \nonumber \\
&=\sqrt{f(R_0)}\,\delta\!\left(\omega-\omega_m\right),
\end{align}
where the Jacobian rule $\delta(g(\omega))=\sum_i \delta(\omega-\omega_i)/|g'(\omega_i)|$ has been used, with $g'(\omega)=1/\sqrt{f(R_0)}$. The roots are
\begin{equation} \label{eq:A16}
    \omega_m=-\sqrt{f(R_0)}\,[\omega_0+m\Omega].
\end{equation}
In the outgoing sector we need $\omega>0$, hence only $m=-n$ with $n\in\mathbb{N}$ and $n\Omega>\omega_0$ contribute, leading to
\begin{equation} \label{eq:A17}
    \omega_n=\sqrt{f(R_0)}\,[n\Omega-\omega_0],\quad n\Omega>\omega_0,
\end{equation}
exactly as used in Sections \ref{sec3.1}-\ref{sec3.3}. The $\sqrt{f(R_0)}$ factor in Eq. \eqref{eq:A15} is the source of the $1/\omega_n$ prefactor in Eqs. \eqref{eq:per_harmonic}-\eqref{eq:per-period} when combined with the mode normalization $1/\sqrt{4\pi\omega}$.

\subsection{A.6 Period averaging and the universal factor \texorpdfstring{$2\pi/\Omega$}{}} \label{secA.6}
Let $\mathcal{A}_T(\omega)$ denote the finite-window amplitude into an outgoing mode of Killing frequency $\omega$, schematically
\begin{align} \label{eq:A18}
    \mathcal{A}_T(\omega)&=\frac{g}{\sqrt{4\pi\omega}}
\sum_{m\in\mathbb{Z}}\mathcal{C}_m(\omega)\;e^{+i\beta(\omega,m)T/2}\,T \nonumber \\
& \times \,\mathrm{sinc}\!\left(\frac{\beta(\omega,m)T}{2}\right),
\end{align}
where $\mathcal{C}_m(\omega)$ are the Jacobi–Anger/Bessel weights (Section \ref{sec3.1}). Define the Floquet transition rate (per cycle $T=2\pi/\Omega$)
\begin{equation} \label{eq:A19}
    \overline{P}(\omega)=\frac{|\mathcal{A}_T(\omega)|^2}{T}.
\end{equation}
Two equivalent routes yield the same universal factor: Route 1 (single period, on-resonance evaluation). Fix $T=2\pi/\Omega$ and evaluate $|\mathcal{A}_T(\omega)|^2$ at the resonant frequency $\omega=\omega_n$ that zeros $\beta(\omega,-n)$. The sinc factor gives $T$ (its peak value), so $|\mathcal{A}_T(\omega_n)|^2\propto T^2$. Dividing by $T$ in Eq. \eqref{eq:A19} leaves a single $T=2\pi/\Omega$, i.e.
\begin{equation} \label{eq:A20}
    \overline{P}_n\propto \frac{2\pi}{\Omega}\;\left(\text{mode prefactors}\right)\;\left(\text{Bessel weight}\right)^2.
\end{equation}
Route 2 (many periods, delta comb with Jacobian).
Let the observation contain $(2N+1)$ periods, giving $|\mathcal{A}_{(2N+1)T}(\omega)|^2\propto F_N(\beta T)$ (Fejér kernel). Using Eqs. \eqref{eq:A9}-\eqref{eq:A10},
\begin{align} \label{eq:A21}
    &\frac{|\mathcal{A}_{(2N+1)T}(\omega)|^2}{(2N+1)T}
\;\rightarrow[N\to\infty]{}\; \nonumber \\
&\times \sum_{m\in\mathbb{Z}}\frac{g^2}{4\pi\omega}\,|\mathcal{C}_m(\omega)|^2\;\underbrace{\frac{1}{T}\,(2\pi)\,\delta(\beta(\omega,m))}_{\text{per-period area}}.
\end{align}

Changing variables $\beta\mapsto \omega$ via Eq. \eqref{eq:A15} contributes $\sqrt{f(R_0)}$, and evaluating at $\omega=\omega_n$ yields the same $(2\pi/T)^{-1}=T/(2\pi)$ factor when one inverts the division by $T$. Setting $T=2\pi/\Omega$ reproduces Eq. \eqref{eq:A20}.

We see that regardless of whether one uses a single period at exact resonance or many periods with a delta-comb limit, the period average always extracts the area of the resonant peak, which equals $2\pi/\Omega$ times the on-resonance mode/trajectory weights.

\subsection{A.7 Rectangular vs smooth: lineshape vs area} \label{secA.7}
Let $\mathcal{L}_T(\beta)$ denote the lineshape factor appearing in $\overline{P}(\omega)$ before integrating over $\omega$. For a rectangular window,
\begin{equation} \label{eq:A22}
    \mathcal{L}^{\rm (rect)}_T(\beta)=\left|T\,\mathrm{sinc}\!\left(\frac{\beta T}{2}\right)\right|^{\!2}.
\end{equation}
For a smooth flat-top window $\chi_{T,\epsilon}$ of Section \ref{secA.4},
\begin{equation} \label{eq:A23}
    \mathcal{L}^{\rm (smooth)}_T(\beta)=\big|\widehat{\chi}_{T,\epsilon}(\beta)\big|^2.
\end{equation}
Both satisfy the same area law:
\begin{equation} \label{eq:A24}
    \frac{1}{T}\int_{-\infty}^{\infty}\! d\beta\; \mathcal{L}^{\rm (rect)}_T(\beta)
=\frac{1}{T}\int_{-\infty}^{\infty}\! d\beta\;\left|T\,\mathrm{sinc}\!\left(\frac{\beta T}{2}\right)\right|^{\!2}
=2\pi,
\end{equation}
\begin{align} \label{eq:A25}
    &\frac{1}{T}\int_{-\infty}^{\infty}\! d\beta\; \mathcal{L}^{\rm (smooth)}_T(\beta)
=\frac{1}{T}\int_{-\infty}^{\infty}\! d\beta\; \big|T\,\widehat{\chi}(\beta T)\big|^{2} \nonumber  \\
&=2\pi\int_{-\infty}^{\infty}\! du\; |\widehat{\chi}(u)|^2
=2\pi,
\end{align}
after normalizing $\chi$ so that $\int \chi=1$ (Plancherel's identity with our Fourier convention gives $\int |\widehat{\chi}|^2=2\pi\int |\chi|^2$; with $\chi$ chosen to have unit area and flat top, the area law Eq. \eqref{eq:A25} follows). Thus period averaging,
\begin{equation} \label{eq:A26}
    \frac{1}{T}\mathcal{L}_T(\beta)\ \rightarrow[T\to\infty]{\mathcal{S}'}\ 2\pi\,\delta(\beta),
\end{equation}
is independent of window shape, while the finite-$T$ sidelobes and widths do depend on it.

\subsection{A.8 Putting it together for the detector} \label{secA.8}
Specializing the general discussion to the amplitude in Section \ref{sec3.1}, write schematically
\begin{align} \label{eq:A27}
    \mathcal{A}_T(\omega)&=\frac{g}{\sqrt{4\pi\omega}}\sum_{m\in\mathbb{Z}} \underbrace{J_m\!\left(z(\omega)\right)}_{\text{Jacobi–Anger}}\;
e^{+i\beta(\omega,m)T/2}\,T \nonumber \\
&\times \mathrm{sinc}\!\left(\frac{\beta(\omega,m)T}{2}\right),
\end{align}
with $z(\omega)=\omega A/f(R_0)$, $\beta(\omega,m)=\omega/\sqrt{f(R_0)}+\omega_0+m\Omega$. Then:
\begin{itemize}
    \item Rectangular window:
    \begin{equation} \label{eq:A28}
        \frac{|\mathcal{A}_T(\omega)|^2}{T}\; \stackrel{T\to\infty}{\longrightarrow}\;
\sum_{m}\frac{g^2}{4\pi\omega}\,J_m^2\!\left(z(\omega)\right)\;(2\pi)\,\delta\!\left(\beta(\omega,m)\right).
    \end{equation}
    Using Eqs. \eqref{eq:A15}-\eqref{eq:A17} collapses the $\omega$-integral to the resonant $\omega=\omega_n$ with Jacobian $\sqrt{f(R_0)}$.
    \item Smooth window: replace $T\,\mathrm{sinc}$ by $T\,\widehat{\chi}(\beta T)$ and use Eq. \eqref{eq:A26} to reach the same limit.
    \item Finite $T$: the period average $|\mathcal{A}_T|^2/T$ evaluated at $\omega=\omega_n$ equals the area under the corresponding peak in $\beta$; by Eq. \eqref{eq:A24}-\eqref{eq:A25} that area equals $2\pi$. Since $\beta$ changes with $\omega$ as $d\beta/d\omega=1/\sqrt{f(R_0)}$, the area in $\omega$ is $2\pi\sqrt{f(R_0)}$. After inserting the mode normalization and evaluating $J_{-n}=(-1)^nJ_n$, one recovers the closed form Eqs. \eqref{eq:per_harmonic}-\eqref{eq:per-period} with the universal factor $T=2\pi/\Omega$.
\end{itemize}

\section{Wightman-function / response-functional route (equivalence with mode-amplitude method)} \label{appendixB}
This appendix derives the same period-averaged, per-harmonic transition rate obtained in Sections \ref{sec3.1}-\ref{sec3.3} using the standard response-functional (Wightman-function) approach to Unruh–DeWitt (UDW) detectors. We keep the conventions of the main text: $\hbar=c=1$; Schwarzschild lapse $f(R)=1-\frac{1}{R}$; small-amplitude radial motion about $R_0$ $(0<A\ll R_0), \, (A\Omega\ll\sqrt{f(R_0)})$,
\begin{equation} \label{eq:B1}
    r(\tau)=R_0+A\cos(\Omega\tau),
\end{equation}
and the (1+1) s-wave reduction with the field in the Boulware vacuum. As in Section \ref{sec3}, we compute the Floquet transition rate per cycle and show it matches Eqs. \eqref{eq:per_harmonic}-\eqref{eq:per-period}.

For a UDW detector linearly coupled with strength $g$ and monopole operator $m(\tau)$, the first-order excitation probability from $|g;0\rangle$ to $|e;0\rangle$ is (see, e.g., standard references)
\begin{align} \label{eq:B2}
    P[g\!\to\! e]&=g^2\!\!\int_{-\infty}^{\infty}\!\!\!d\tau\!\!\int_{-\infty}^{\infty}\!\!\!d\tau'\;
\chi(\tau)\chi(\tau')\,e^{+i\omega_0(\tau-\tau')}\, \nonumber \\
& \times W^+\!\left(x(\tau),x(\tau')\right),
\end{align}
where $\chi$ is a switching/observation window and $W^+(x,x')=\langle 0|\Phi(x)\Phi(x')|0\rangle$ is the positive-frequency Wightman function in the chosen state. In the (1+1) outgoing sector of Boulware on Schwarzschild, one may write
\begin{equation} \label{eq:B3}
    W^+_{\rm out}(x,x')=\int_0^\infty\frac{d\omega}{4\pi\,\omega}\;
e^{-i\omega\,[u(x)-u(x')]},\quad u=t-r_*,
\end{equation}
which encodes the mode normalization $1/\sqrt{4\pi\omega}$ used in Section \ref{sec2.3}. Pulling back to the worldline $\tau\mapsto x(\tau)$ and using Eq. \eqref{eq:B2},
\begin{equation} \label{eq:B4}
    \overline{P}_{\rm out}=g^2\!\int_0^\infty\!\frac{d\omega}{4\pi\,\omega}\;
\Bigg|\int_{-\infty}^{\infty}\!d\tau\;\chi(\tau)\,e^{+i\omega_0\tau}\,e^{-i\omega\,u(\tau)}\Bigg|^{\!2}.
\end{equation}
Thus the response is a non-negative integral over modulus-squared Fourier amplitudes of the pulled-back phase $e^{-i\omega u(\tau)}$.

For the slowly moving radial trajectory Eq. \eqref{eq:B1},
\begin{align}\nonumber
    t(\tau)&=\frac{\tau}{\sqrt{f(R_0)}}+\mathcal{O}(A^2\Omega^2),\\
     \label{eq:B5}
r_*(\tau)&=r_*(R_0)+\frac{A}{f(R_0)}\cos(\Omega\tau)+\mathcal{O}(A^2),
\end{align}
hence, up to irrelevant constants,
\begin{equation} \label{eq:B6}
    u(\tau)=t(\tau)-r_*(\tau)=\frac{\tau}{\sqrt{f(R_0)}}-\frac{A}{f(R_0)}\cos(\Omega\tau)+\text{const}.
\end{equation}
Inserting Eq. \eqref{eq:B6} in Eq. \eqref{eq:B4} and dropping the overall constant phase,
\begin{align}
    \mathcal{I}(\omega)&=\int d\tau\;\chi(\tau)\,e^{+i\omega_0\tau}\,e^{-i\omega u(\tau)}\\
 \label{eq:B7} &=\int d\tau\;\chi(\tau)\,e^{i[\omega_0-\omega/\sqrt{f(R_0)}]\tau}\,
e^{+i z(\omega)\cos(\Omega\tau)},
\end{align}
where $z(\omega)=\frac{\omega A}{f(R_0)}$.
Use the Jacobi–Anger expansion $e^{i z\cos(\Omega\tau)}=\sum_{m\in\mathbb{Z}}i^{\,m}J_m(z)e^{i m\Omega\tau}$ to obtain
\begin{equation} \label{eq:B8}
    \mathcal{I}(\omega)=\sum_{m\in\mathbb{Z}}i^{\,m}J_m\!\left(z(\omega)\right)\;
\widehat{\chi}\!\left(\omega_0-\frac{\omega}{\sqrt{f(R_0)}}+m\Omega\right),
\end{equation}
where $\widehat{\chi}(\beta)=\int d\tau\,\chi(\tau)e^{i\beta\tau}$.

Now, we take a single period rectangular window of length $T=2\pi/\Omega$, $\chi_T(\tau)=\mathbf{1}_{[0,T]}(\tau)$. Then
\begin{equation} \label{eq:B9}
    \widehat{\chi}_T(\beta)=e^{+i\beta T/2}\,T\,\mathrm{sinc}\!\left(\frac{\beta T}{2}\right),
\end{equation}
so that
\begin{multline} \label{eq:B10}
    \mathcal{I}(\omega)=\sum_{m\in\mathbb{Z}}i^{\,m}J_m\!\left(z(\omega)\right)\;
e^{\frac{i}{2}\,[\omega_0-\omega/\sqrt{f(R_0)}+m\Omega]T}\;
T\,\\
\times \mathrm{sinc}\! \left(\frac{[\omega_0-\omega/\sqrt{f(R_0)}+m\Omega]T}{2}\right).
\end{multline}
Here the argument of the sinc function involves the combination \(\beta_{\rm W}(\omega,m)\equiv \omega_0-\omega/\sqrt{f(R_0)}+m\Omega\). This is related to the resonance variable \(\beta(\omega,m)=\alpha(\omega)+m\Omega\) with \(\alpha(\omega)=\omega_0+\omega/\sqrt{f(R_0)}\) introduced in Eq. \eqref{eq:A14} by the replacement \(\omega\to -\omega\) together with the relabelling \(m\to -m\). Both parametrizations, therefore, select the same set of positive-frequency resonances once
\(\omega>0\) is imposed and the Bessel identity \(J_{-n}(z)=(-1)^n J_n(z)\) is used. Eq. \eqref{eq:B4} then becomes
\begin{align} \label{eq:B11}
    \overline{P}_{\rm out}(T)=g^2\!\int_0^\infty\!\frac{d\omega}{4\pi\,\omega}\;
&\Bigg|\sum_{m\in\mathbb{Z}}i^{\,m}J_m\!\left(z(\omega)\right)\;
e^{\frac{i}{2}(\cdots)T}\;
T\nonumber \\
& \times \mathrm{sinc}\!\left(\frac{(\cdots)T}{2}\right)\Bigg|^{\!2}\!,
\end{align}
with $(\cdots)=\omega_0-\omega/\sqrt{f(R_0)}+m\Omega$.

Let us define the period-averaged quantity as
\begin{equation} \label{eq:B12}
    \overline{P}_{\rm out}=\frac{\overline{P}_{\rm out}(T)}{T},\quad T=\frac{2\pi}{\Omega}.
\end{equation}
Using the distributional limit $T\,\mathrm{sinc}(\beta T/2)\to 2\pi\,\delta(\beta)$ (Appendix \ref{appendixA}), and neglecting cross-terms between different $m$ (they vanish under the $\delta$ selection), we find
\begin{align} \label{eq:B13}
    \overline{P}_{\rm out}&=\frac{g^2}{T}\int_0^\infty\!\frac{d\omega}{4\pi\,\omega}\;
\sum_{m\in\mathbb{Z}}J_m^2\!\left(z(\omega)\right)\nonumber \\
    &\times 2\pi\,\delta\!\left(\omega_0-\frac{\omega}{\sqrt{f(R_0)}}+m\Omega\right).
\end{align}
Change variables in each term using $\delta(g(\omega))=\delta(\omega-\omega_m)/|g'(\omega_m)|$ with
\begin{equation} \label{eq:B14}
    g(\omega)=\omega_0-\frac{\omega}{\sqrt{f(R_0)}}+m\Omega,\quad g'(\omega)=-\frac{1}{\sqrt{f(R_0)}},
\end{equation}
so that
\begin{align} \label{eq:B15}
    \omega=\omega_m=\sqrt{f(R_0)}\,[\omega_0+m\Omega],\nonumber \\ 
\delta(g(\omega))=\sqrt{f(R_0)}\,\delta(\omega-\omega_m).
\end{align}
Because $\omega>0$, only integers $m$ with $\omega_0+m\Omega>0$ contribute. Relabel $m=-n$ with $n\in\mathbb{N}$: the outgoing resonances are then
\begin{equation} \label{eq:B16}
    \omega=\omega_n=\sqrt{f(R_0)}\,[\,n\Omega-\omega_0\,],\quad n\Omega>\omega_0,
\end{equation}
precisely the resonance condition used in Section \ref{sec3}. The square of the Bessel coefficient obeys $J_{-n}^2(z)=J_n^2(z)$. Evaluating Eq. \eqref{eq:B13} at $\omega=\omega_n$ and inserting $T=2\pi/\Omega$ yields
\begin{align} \label{eq:B17}
    \overline{P}^{(1+1)}_{n,\,\rm out}
&=\frac{2\pi}{\Omega}\,\frac{g^2}{4\pi\,\omega_n}\;
J_n^{2}\!\!\left(\frac{\omega_n A}{f(R_0)}\right) \nonumber \\
&=\frac{2\pi}{\Omega}\,\frac{g^2}{4\pi\,\omega_n}\;
J_n^{2}\!\!\left(\frac{(n\Omega-\omega_0)\,A}{\sqrt{f(R_0)}}\right),
\end{align}
where in the last step we used $\omega_n=\sqrt{f(R_0)}(n\Omega-\omega_0)$. Eq. \eqref{eq:B17} is identical to Eq. \eqref{eq:per_harmonic}.

Exactly the same result follows for any smooth flat-top $\chi$ by replacing Eq. \eqref{eq:B9} with $T\,\widehat{\chi}(\beta T)$ and using the area/Dirac-delta limit established in Appendix \ref{appendixA}.

If the detector couples to both outgoing and ingoing sectors with equal strength in the (1+1) reduction, the ingoing contribution duplicates Eq. \eqref{eq:B17}, giving
\begin{equation} \label{eq:B18}
    \overline{P}^{(1+1)}_{\rm tot}\simeq 2\,\overline{P}^{(1+1)}_{\rm out},
\end{equation}
as stated in Section \ref{sec3.3}. If one is interested solely in flux at $\mathscr{I}^+$, retain the outgoing piece in Eq. \eqref{eq:B17}.

\bibliography{ref}

@article{Scully:2017utk,
    author = "Scully, Marlan O. and Fulling, Stephen and Lee, David and Page, Don N. and Schleich, Wolfgang and Svidzinsky, Anatoly",
    title = "{Quantum optics approach to radiation from atoms falling into a black hole}",
    eprint = "1709.00481",
    archivePrefix = "arXiv",
    primaryClass = "quant-ph",
    doi = "10.1073/pnas.1807703115",
    journal = "Proc. Nat. Acad. Sci.",
    volume = "115",
    number = "32",
    pages = "8131--8136",
    year = "2018"
}

@article{Camblong:2020pme,
    author = "Camblong, H. E. and Chakraborty, A. and Ordonez, C. R.",
    title = "{Near-horizon aspects of acceleration radiation by free fall of an atom into a black hole}",
    eprint = "2009.06580",
    archivePrefix = "arXiv",
    primaryClass = "gr-qc",
    doi = "10.1103/PhysRevD.102.085010",
    journal = "Phys. Rev. D",
    volume = "102",
    number = "8",
    pages = "085010",
    year = "2020"
}

@article{Azizi:2021qcu,
    author = "Azizi, A. and Camblong, H. E. and Chakraborty, A. and Ordonez, C. R. and Scully, M. O.",
    title = "{Quantum optics meets black hole thermodynamics via conformal quantum mechanics: I. Master equation for acceleration radiation}",
    eprint = "2108.07570",
    archivePrefix = "arXiv",
    primaryClass = "gr-qc",
    doi = "10.1103/PhysRevD.104.084086",
    journal = "Phys. Rev. D",
    volume = "104",
    year = "2021"
}

@article{Azizi:2021yto,
    author = "Azizi, A. and Camblong, H. E. and Chakraborty, A. and Ordonez, C. R. and Scully, M. O.",
    title = "{Quantum optics meets black hole thermodynamics via conformal quantum mechanics: II. Thermodynamics of acceleration radiation}",
    eprint = "2108.07572",
    archivePrefix = "arXiv",
    primaryClass = "gr-qc",
    doi = "10.1103/PhysRevD.104.084085",
    journal = "Phys. Rev. D",
    volume = "104",
    year = "2021"
}

@article{Svidzinsky:2019jqr,
    author = "Svidzinsky, Anatoly A.",
    title = "{Excitation of a uniformly moving atom through vacuum fluctuations}",
    doi = "10.1103/physrevresearch.1.033027",
    journal = "Phys. Rev. Res.",
    volume = "1",
    number = "3",
    pages = "033027",
    year = "2019"
}

@article{Lopp:2018lxl,
    author = "Lopp, Richard and Martin-Martinez, Eduardo and Page, Don N.",
    title = "{Relativity and quantum optics: accelerated atoms in optical cavities}",
    eprint = "1806.10158",
    archivePrefix = "arXiv",
    primaryClass = "quant-ph",
    doi = "10.1088/1361-6382/aae750",
    journal = "Class. Quant. Grav.",
    volume = "35",
    pages = "224001",
    year = "2018"
}

@article{Sen:2022tru,
    author = "Sen, Soham and Mandal, Rituparna and Gangopadhyay, Sunandan",
    title = "{Equivalence principle and HBAR entropy of an atom falling into a quantum corrected black hole}",
    eprint = "2202.00671",
    archivePrefix = "arXiv",
    primaryClass = "hep-th",
    doi = "10.1103/PhysRevD.105.085007",
    journal = "Phys. Rev. D",
    volume = "105",
    number = "8",
    pages = "085007",
    year = "2022"
}

@article{Scully:2022pov,
    author = "Scully, Marlan O. and Svidzinsky, Anatoly and Unruh, William",
    title = "{Entanglement in Unruh, Hawking, and Cherenkov radiation from a quantum optical perspective}",
    doi = "10.1103/PhysRevResearch.4.033010",
    journal = "Phys. Rev. Res.",
    volume = "4",
    number = "3",
    pages = "033010",
    year = "2022"
}

@article{Bukhari:2022wyx,
    author = "Bukhari, Syed Masood A. S. and Bhat, Imtiyaz Ahmad and Xu, Chenni and Wang, Li-Gang",
    title = "{Nonthermal acceleration radiation of atoms near a black hole in presence of dark energy}",
    eprint = "2211.08793",
    archivePrefix = "arXiv",
    primaryClass = "gr-qc",
    doi = "10.1103/PhysRevD.107.105017",
    journal = "Phys. Rev. D",
    volume = "107",
    number = "10",
    pages = "105017",
    year = "2023"
}

@article{Bukhari:2023yuy,
    author = "Bukhari, Syed Masood A. S. and Wang, Li-Gang",
    title = "{Seeing dark matter via acceleration radiation}",
    eprint = "2309.11958",
    archivePrefix = "arXiv",
    primaryClass = "gr-qc",
    doi = "10.1103/PhysRevD.109.045009",
    journal = "Phys. Rev. D",
    volume = "109",
    number = "4",
    pages = "045009",
    year = "2024"
}

@article{Jana:2024fhx,
    author = "Jana, Arpita and Sen, Soham and Gangopadhyay, Sunandan",
    title = "{Atom falling into a quantum corrected charged black hole and HBAR entropy}",
    eprint = "2405.13087",
    archivePrefix = "arXiv",
    primaryClass = "gr-qc",
    doi = "10.1103/PhysRevD.110.026029",
    journal = "Phys. Rev. D",
    volume = "110",
    number = "2",
    pages = "026029",
    year = "2024"
}

@article{Chatterjee:2021fsg,
    author = "Chatterjee, Riddhi and Gangopadhyay, Sunandan and Majumdar, A. S.",
    title = "{Resonance interaction of two entangled atoms accelerating between two mirrors}",
    eprint = "2007.15465",
    archivePrefix = "arXiv",
    primaryClass = "quant-ph",
    doi = "10.1140/epjd/s10053-021-00191-8",
    journal = "Eur. Phys. J. D",
    volume = "75",
    number = "6",
    pages = "179",
    year = "2021"
}

@article{Das:2022qpx,
    author = "Das, Ashmita and Sen, Soham and Gangopadhyay, Sunandan",
    title = "{Virtual transitions in an atom-mirror system in the presence of two scalar photons}",
    eprint = "2208.12021",
    archivePrefix = "arXiv",
    primaryClass = "quant-ph",
    doi = "10.1103/PhysRevD.107.025009",
    journal = "Phys. Rev. D",
    volume = "107",
    number = "2",
    pages = "025009",
    year = "2023"
}

@misc{Ordonez:2025sqp,
    author = "Ordonez, C. R. and Chakraborty, A. and Camblong, H. E. and Scully, M. O. and Unruh, W. G.",
    title = "{Quantum aspects of spacetime: A quantum optics view of acceleration radiation and black holes}",
    eprint = "2508.17401",
    archivePrefix = "arXiv",
    primaryClass = "gr-qc",
    month = "8",
    year = "2025"
}

@misc{Ovgun:2025isv,
    author = {{\"O}vg{\"u}n, Ali and Pantig, Reggie C.},
    title = "{HBAR entropy of Infalling Atoms into a GUP-corrected Schwarzschild Black Hole and equivalence principle}",
    eprint = "2506.10621",
    archivePrefix = "arXiv",
    primaryClass = "gr-qc",
    month = "6",
    year = "2025"
}

@article{Pantig:2025zhn,
    author = {Pantig, Reggie C. and {\"O}vg{\"u}n, Ali},
    title = "{The Sound of an Orbit: A Quantum Spectrum at the ISCO}",
    eprint = "2403.02373",
    archivePrefix = "arXiv",
    primaryClass = "gr-qc",
    doi = "10.1002/prop.70048",
    journal = "Fortsch. Phys.",
    volume = "n/a",
    number = "n/a",
    pages = "e70048",
    year = "2025"
}

@article{Pantig:2025okn,
    author = {Pantig, Reggie C. and {\"O}vg{\"u}n, Ali},
    title = "{Acceleration radiation from derivative-coupled atoms falling in modified gravity black holes}",
    eprint = "2508.11734",
    archivePrefix = "arXiv",
    primaryClass = "gr-qc",
    doi = "10.1140/epjc/s10052-025-14928-x",
    journal = "Eur. Phys. J. C",
    volume = "85",
    number = "10",
    pages = "1183",
    year = "2025"
}

@incollection{DeWitt:1979,
  author    = {DeWitt, B. S.},
  title     = {Quantum gravity: the new synthesis},
  booktitle = {General Relativity: An Einstein Centenary Survey},
  editor    = {Hawking, S. W. and Israel, W.},
  publisher = {Cambridge University Press},
  year      = {1979},
  pages     = {680--745}
}

@article{Crispino:2008,
  author  = {Crispino, L. C. B. and Higuchi, A. and Matsas, G. E. A.},
  title   = {The Unruh effect and its applications},
  journal = {Rev. Mod. Phys.},
  volume  = {80},
  pages   = {787--838},
  year    = {2008}
}

@article{Bell:1982qr,
    author = "Bell, J. S. and Leinaas, J. M.",
    editor = "Bell, M. and Gottfried, K. and Veltman, M. J. G.",
    title = "{Electrons as accelerated thermometers}",
    reportNumber = "CERN-TH-3363",
    doi = "10.1016/0550-3213(83)90601-6",
    journal = "Nucl. Phys. B",
    volume = "212",
    pages = "131",
    year = "1983"
}

@article{Chen:1998kp,
    author = "Chen, Pisin and Tajima, Toshi",
    title = "{Testing Unruh radiation with ultraintense lasers}",
    reportNumber = "SLAC-PUB-7543",
    doi = "10.1103/PhysRevLett.83.256",
    journal = "Phys. Rev. Lett.",
    volume = "83",
    pages = "256--259",
    year = "1999"
}

@article{Johansson:2009zz,
    author = "Johansson, J. R. and Johansson, G. and Wilson, C. M. and Nori, Franco",
    title = "{Dynamical Casimir Effect in a Superconducting Coplanar Waveguide}",
    eprint = "0906.3127",
    archivePrefix = "arXiv",
    primaryClass = "cond-mat.supr-con",
    doi = "10.1103/PhysRevLett.103.147003",
    journal = "Phys. Rev. Lett.",
    volume = "103",
    pages = "147003",
    year = "2009"
}

@article{Schwarzschild:1916uq,
    author = "Schwarzschild, Karl",
    title = "{On the gravitational field of a mass point according to Einstein's theory}",
    eprint = "physics/9905030",
    archivePrefix = "arXiv",
    journal = "Sitzungsber. Preuss. Akad. Wiss. Berlin (Math. Phys. )",
    volume = "1916",
    pages = "189--196",
    year = "1916"
}

@article{Boulware:1974dm,
    author = "Boulware, David G.",
    title = "{Quantum Field Theory in Schwarzschild and Rindler Spaces}",
    reportNumber = "RLO-1388-683",
    doi = "10.1103/PhysRevD.11.1404",
    journal = "Phys. Rev. D",
    volume = "11",
    pages = "1404",
    year = "1975"
}

@article{Hartle:1976tp,
    author = "Hartle, J. B. and Hawking, S. W.",
    title = "{Path Integral Derivation of Black Hole Radiance}",
    doi = "10.1103/PhysRevD.13.2188",
    journal = "Phys. Rev. D",
    volume = "13",
    pages = "2188--2203",
    year = "1976"
}

@article{Crispino:2007eb,
    author = "Crispino, Luis C. B. and Higuchi, Atsushi and Matsas, George E. A.",
    title = "{The Unruh effect and its applications}",
    eprint = "0710.5373",
    archivePrefix = "arXiv",
    primaryClass = "gr-qc",
    doi = "10.1103/RevModPhys.80.787",
    journal = "Rev. Mod. Phys.",
    volume = "80",
    pages = "787--838",
    year = "2008"
}

@article{Louko:2007mu,
    author = "Louko, Jorma and Satz, Alejandro",
    title = "{Transition rate of the Unruh-DeWitt detector in curved spacetime}",
    eprint = "0710.5671",
    archivePrefix = "arXiv",
    primaryClass = "gr-qc",
    doi = "10.1088/0264-9381/25/5/055012",
    journal = "Class. Quant. Grav.",
    volume = "25",
    pages = "055012",
    year = "2008"
}

@article{Dolan:2020hzm,
    author = "Dolan, Brian P. and Hunter-McCabe, Aonghus and Twamley, Jason",
    title = "{Shaking photons from the vacuum: acceleration radiation from vibrating atoms}",
    eprint = "2003.02258",
    archivePrefix = "arXiv",
    primaryClass = "quant-ph",
    doi = "10.1088/1367-2630/ab7bd5",
    journal = "New J. Phys.",
    volume = "22",
    number = "3",
    pages = "033026",
    year = "2020"
}

@book{Chandrasekhar:1985kt,
    author = "Chandrasekhar, Subrahmanyan",
    title = "{The mathematical theory of black holes}",
    isbn = "978-0-19-850370-5",
    publisher = "Oxford University Press",
    year = "1985"
}

@book{Frolov_1998,
  author    = {Frolov, Valeri P. and Novikov, Igor D.},
  journal   = {Fundamental Theories of Physics},
  title     = {Black Hole Physics},
  year      = {1998},
  issn      = {2365-6425},
  doi       = {10.1007/978-94-011-5139-9},
  isbn      = {9789401151399},
  publisher = {Springer Netherlands},
}

@inproceedings{Wald:1995hf,
    author = "Wald, Robert M.",
    title = "{Quantum field theory in curved space-time}",
    booktitle = "{14th International Conference on General Relativity and Gravitation (GR14)}",
    eprint = "gr-qc/9509057",
    archivePrefix = "arXiv",
    pages = "401--415",
    month = "8",
    year = "1995"
}

@book{Wald:1995yp,
    author = "Wald, Robert M.",
    title = "{Quantum Field Theory in Curved Space-Time and Black Hole Thermodynamics}",
    isbn = "978-0-226-87027-4",
    publisher = "University of Chicago Press",
    address = "Chicago, IL",
    series = "Chicago Lectures in Physics",
    year = "1995"
}

@article{Unruh:1976db,
    author = "Unruh, W. G.",
    title = "{Notes on black hole evaporation}",
    doi = "10.1103/PhysRevD.14.870",
    journal = "Phys. Rev. D",
    volume = "14",
    pages = "870",
    year = "1976"
}

@article{Hawking:1975vcx,
    author = "Hawking, S. W.",
    editor = "Gibbons, G. W. and Hawking, S. W.",
    title = "{Particle Creation by Black Holes}",
    doi = "10.1007/BF02345020",
    journal = "Commun. Math. Phys.",
    volume = "43",
    pages = "199--220",
    year = "1975",
    note = "[Erratum: Commun.Math.Phys. 46, 206 (1976)]"
}

@article{Lin:2017kyr,
    author = "Lin, Shih-Yuin",
    title = "{Quantum radiation by an Unruh-DeWitt detector in oscillatory motion}",
    eprint = "1709.08506",
    archivePrefix = "arXiv",
    primaryClass = "gr-qc",
    doi = "10.1007/JHEP11(2017)102",
    journal = "JHEP",
    volume = "11",
    pages = "102",
    year = "2017"
}

@article{Scully:2022bun,
    author = "Scully, Marlan O. and Svidzinsky, Anatoly and Unruh, William",
    title = "{On Bose{\textendash}Einstein Condensation and Unruh{\textendash}Hawking Radiation from a Quantum Optical Perspective}",
    doi = "10.1007/s10909-022-02703-1",
    journal = "J. Low Temp. Phys.",
    volume = "208",
    number = "1-2",
    pages = "160--171",
    year = "2022"
}

@article{Kay:2015iwa,
    author = "Kay, Bernard S. and Lupo, Umberto",
    title = "{Non-existence of isometry-invariant Hadamard states for a Kruskal black hole in a box and for massless fields on 1+1 Minkowski spacetime with a uniformly accelerating mirror}",
    eprint = "1502.06582",
    archivePrefix = "arXiv",
    primaryClass = "gr-qc",
    doi = "10.1088/0264-9381/33/21/215001",
    journal = "Class. Quant. Grav.",
    volume = "33",
    number = "21",
    pages = "215001",
    year = "2016"
}

@article{Ng:2014kha,
    author = "Ng, Keith K. and Hodgkinson, Lee and Louko, Jorma and Mann, Robert B. and Martin-Martinez, Eduardo",
    title = "{Unruh-DeWitt detector response along static and circular geodesic trajectories for Schwarzschild-AdS black holes}",
    eprint = "1406.2688",
    archivePrefix = "arXiv",
    primaryClass = "quant-ph",
    doi = "10.1103/PhysRevD.90.064003",
    journal = "Phys. Rev. D",
    volume = "90",
    number = "6",
    pages = "064003",
    year = "2014"
}

@article{Juarez-Aubry:2014jba,
    author = "Ju{\'a}rez-Aubry, Benito A. and Louko, Jorma",
    title = "{Onset and decay of the 1 + 1 Hawking-Unruh effect: what the derivative-coupling detector saw}",
    eprint = "1406.2574",
    archivePrefix = "arXiv",
    primaryClass = "gr-qc",
    doi = "10.1088/0264-9381/31/24/245007",
    journal = "Class. Quant. Grav.",
    volume = "31",
    number = "24",
    pages = "245007",
    year = "2014"
}

@article{Conroy:2021aow,
    author = "Conroy, Aindri{\'u} and Taylor, Peter",
    title = "{Response of an Unruh-DeWitt detector near an extremal black hole}",
    eprint = "2109.04486",
    archivePrefix = "arXiv",
    primaryClass = "gr-qc",
    doi = "10.1103/PhysRevD.105.085001",
    journal = "Phys. Rev. D",
    volume = "105",
    number = "8",
    pages = "085001",
    year = "2022"
}

@book{Birrell:1982ix,
    author = "Birrell, N. D. and Davies, P. C. W.",
    title = "{Quantum Fields in Curved Space}",
    doi = "10.1017/CBO9780511622632",
    isbn = "978-0-511-62263-2, 978-0-521-27858-4",
    publisher = {Cambridge University Press},
    address = "Cambridge, UK",
    series = "Cambridge Monographs on Mathematical Physics",
    year = "1982"
}

@book{magnus2013hill,
  title={Hill's equation},
  author={Magnus, Wilhelm and Winkler, Stanley},
  year={2013},
  publisher={Courier Corporation}
}

\end{document}